\documentclass[a4paper,11pt]{article}
\pdfoutput=1
\usepackage[utf8x]{inputenc}
\usepackage{jheppub}

\usepackage[T1]{fontenc} 

\usepackage{amstext,amssymb}
\usepackage{amsmath}
\usepackage{graphicx}
\usepackage{xspace}
\usepackage{color}
\usepackage{units}
\usepackage{multirow}
\usepackage{slashed} 
\usepackage{comment}
\usepackage{subfigure}
\usepackage[]{hyperref}
\usepackage{appendix}
\usepackage{url}

\newcommand{\bea}{\begin{eqnarray}}
\newcommand{\eea}{\end{eqnarray}}

\title{\boldmath Vacuum Structure of Alternative Left-Right Model}
\author[a]{\small \hspace*{-0.5cm} Mariana Frank}
\author[b]{\small , Chayan Majumdar}
\author[c]{\small , Poulose Poulose}
\author[b]{\small , Supriya Senapati}
\author[b]{\small , Urjit A. Yajnik}
\affiliation[a]{Department of Physics,  Concordia University, \\
7141 Sherbrooke St. West, Montreal, Quebec, Canada H4B 1R6}
\affiliation[b]{Department of Physics, Indian Institute of Technology Bombay, \\
Powai, Mumbai, Maharashtra 400 076, India}
\affiliation[c]{Department of Physics, Indian Institute of Technology Guwahati, \\
Assam 781 039, India}
\emailAdd{mariana.frank@concordia.ca}
\emailAdd{chayan@phy.iitb.ac.in}
\emailAdd{poulose@iitg.ac.in}
\emailAdd{supriya@phy.iitb.ac.in}
\emailAdd{yajnik@iitb.ac.in}
\abstract{
We investigate an Alternative Left-Right Model (ALRM) with $SU(2)_L$ as well as $SU(2)_R$ gauge groups, but unlike the traditional left-right symmetric models (LRSM) is not symmetric under the exchange of the fermion content. Interestingly, it can be embedded in $E_6$, while its low energy Higgs structure resembles the LRSM, involving Higgs doublets $\chi_{L,R}$ and one Higgs bidoublet $\Phi$. We analyze the scalar potential and the vacuum structure of the theory analytically to ensure the stability of scalar potential  via bounded from below (BFB) and copositivity criteria, accompanied by a numerical study. We establish the necessary criteria for electric charge-preserving vacua, yielding constraints on various coupling parameters of the theory. Finally we obtain constraints on the parameters of the model from collider data on the masses of the Higgs scalars.}

\begin{document} 
\maketitle
\flushbottom
\section{Introduction}
\label{sec:intro}
The Standard Model (SM) of particle physics is very successful in describing the properties of elementary particles and their interactions. The theoretical predictions of the model match collider searches with high accuracy. Based on the gauge group $SU(3)_C \otimes SU(2)_L \otimes U(1)_Y$ it provides a successful description of the strong and the electroweak phenomena. However, several notable phenomena cannot be addressed within this SM framework. Being a chiral framework, it admits maximal violation of parity in low-energy weak interactions, although the fermionic content subject only to the strong and electromagnetic forces, while chiral, turns out to be parity balanced. While introduced in a phenomenological manner in the SM,  the origin of parity violation is expected to arise naturally in a more fundamental description at higher energies.  Also, due to absence of a right-handed partner of the left-handed neutrinos, SM cannot explain the  observations of neutrino experiments like SuperK \cite{Super-Kamiokande:2001bfk}, SNO  \cite{SNO:2002tuh}, which require that active neutrinos have small but finite mass. Other non-negotiable situations to go beyond the SM, connected to cosmology, are the large presence of Dark Matter in the Universe and the observed baryon asymmetry of the Universe (BAU).  While the Higgs mechanism for electroweak symmetry breaking is established by observations at the LHC, there are difficulties here. One subtle issue is related to the structure of vacuum.  What we know from current experimental observations of the relevant parameters is that the vacuum as described within the SM is not absolutely stable, with possibility of the Higgs quartic coupling turning negative at high energies \cite{Isidori:2001bm}. 
These puzzles definitely indicate that  SM, while describing particle dynamics very well at the electroweak scale, require modifications at higher energies.

One of the phenomenologically most attractive scenarios of Beyond the Standard Model  (BSM) perspective is the Left-Right Symmetric Model (LRSM) \cite{PhysRevD.11.2558, PhysRevD.10.275, PhysRevD.12.1502, Senjanovic:1978ev, PhysRevLett.44.912, PhysRevD.23.165, PhysRevD.8.1240, PhysRevLett.32.1083}. Based on the extended gauge structure $SU(3)_C \otimes SU(2)_L \otimes SU(2)_R \otimes U(1)_{B-L}$, this model can address some of the above mentioned inconsistencies. Left-Right models  relate the maximal breaking of parity in the SM with the origin of smallness of light  neutrino masses. One can think of SM as a low-energy effective field theory of such LRSMs. Unlike in the SM, right-handed neutrinos are naturally included in right-handed doublets within this framework leading to light neutrino mass via the so-called seesaw mechanism \cite{Mohapatra:1974gc, Mohapatra:1977mj}.  In the scalar sector, the Higgs doublet present in the SM is elevated to a bidoublet so as to connect between the left- and the right-sectors. In addition, the breakdown of  LRSM gauge symmetry to the SM gauge symmetry requires the presence of additional scalar fields.  There are different possibilities here, including  (i) one Higgs bidoublet $\Phi$ and two scalar doublets $\chi_{L,R}$ \cite{Majumdar:2018eqz, Senapati:2020alx, FileviezPerez:2016erl}, (ii) one Higgs bidoublet and two triplets $\Delta_{L,R}$ \cite{PhysRevLett.44.912} or (iii) combination of the above two. 
 In LRSM the baryon number  ($B$) and the lepton number ($L$) enter as the generator of $U(1)_{B-L}$ gauge symmetry, with the left-right symmetry breaking leading to $SU(2)_R\otimes U(1)_{B-L}\to U(1)_Y$, where $Y$ is the hypercharge of the SM.

An interesting feature of the left-right models is that they can be easily incorporated into a larger group such as $SO(10), ~SU(5)$ or $E_6$, which can unify the electroweak and the strong interaction in the context of grand unified scenario (GUTs) \cite{PhysRevD.10.275, PhysRevLett.32.438, PhysRevLett.33.451, Fritzsch:1974nn}. LRSM, while quite successful as a BSM scenario, unfortunately  suffers from unavoidable flavor-changing neutral current (FCNC) processes \cite{Ecker:1983uh, Zhang:2007da, Maiezza:2010ic}, mediated by the additional neutral Higgs bosons present, consequently requiring them to be very heavy to avoid the present constraints.
However, an alternative version can be found, broadly still within the LRSM class, in which the large FCNC are avoided. It was known that the gauge group considered for LRSM could be embedded into the exceptional group $E_6$ (see \cite{Hewett:1988xc, Langacker:1998tc} for reviews).   $E_6$  has two maximal subgroup structures, $SO(10) \otimes U(1)$ and $SU(3) \otimes SU(3) \otimes SU(3)$. The usual or canonical LRSM described above can be embedded in both of these subgroups, with the $\textbf{16}$ representation of $SO(10)$ incorporating the whole fermion spectrum of the mode.  Under  the $SU(3) \otimes SU(3) \otimes SU(3)$ these are distributed in the $(\textbf{3},~\textbf{3}^*,~\textbf{1}),~ (\textbf{3}^*,~\textbf{3},~\textbf{1}),~(\textbf{1},~\textbf{3},~\textbf{3}^*)$ representations.  Specifically, Ernest Ma \cite{PhysRevD.36.274, PhysRevD.36.878}  noted that it is possible to assign the fermion content of the Left-Right model partially to the \textbf{16} and partially to the \textbf{10} representations of the $SO(10)$, leading to a novel set up where the $W_R$ no longer interact with the right-handed down-type quarks, and further, avoid the dangerous tree-level FCNC interactions. In the $SU(3) \otimes SU(3) \otimes SU(3)$ representations, this corresponds to rearrangement of the rows of $(\textbf{3}^*,~\textbf{3},~\textbf{1})$ and columns of $(\textbf{1},~\textbf{3},~\textbf{3}^*)$  \cite{Ma:2010us, Frank:2019nid}, and consequently to a rearrangement of the quark and lepton multiplets of its subgroup $SU(3) \otimes SU(2) \otimes SU(2)\otimes U(1)$, which is considered as the gauge group of the usual LRSM.

Such alternative scenarios of the  left-right models  are known as Alternative Left-Right Models (ALRMs). Such models remain chiral, and can be considered to be balanced in containing both the left and right  chiralities of the fermions, however do not obey any left-right exchange symmetry. The right handed (RH) currents of the traditional LRSM remain inaccessible to colliders due to the concomitant constraint of preventing FCNC, but the ALRM do not face the latter constraint and permit an accessible RH breaking scale of a few TeV, making ALRMs more interesting from the perspective of collider phenomenology. Further, the conventional non-supersymmetric version of the LRSM model, while addressing the neutrino mass generation in a natural way, does not include a potential DM candidate. On the other hand, in the ALRMs one of the neutral fermions could be a natural scotino stabilised through an additional global symmetry ($S$) imposed in the scenario called Dark Left-Right Model (DLRM) \cite{Khalil:2009nb, Khalil:2010yt, Ma:2010us}. 
Having different assignments of the $S$-charge the scotino could be a Majorana fermion (DLRM1) \cite{Khalil:2009nb} or a Dirac fermion (DLRM2)  \cite{Khalil:2010yt}, with the former scenario allowing to define a generalised lepton number $L=S-T_{3R}$, whereas the latter defines $L=S+T_{3R}$.  A third possibility, which we shall call DLRM3, discussed in Ref.\cite{Ma:2012gb} is compatible with very light (in the keV range) Majorana scotino dark matter as against heavy scotinos (in the $10^2$ GeV range) in the earlier two cases.  Further, more recently, a gauged $U(1)_S$ possibility was considered with additional fermions incorporated to render the model anomaly free \cite{Kownacki:2017sqn}, consequently providing multiple dark matter candidates. Collider phenomenology focusing on the heavy gauge bosons was studied in the context of DLRM2 by Refs.\cite{Khalil:2010yt, Frank:2019nid}, with the latter in addition performing a detailed study of its dark matter sector exploring viable parameter space regions. DLRM3 was considered in detail by Refs.\cite{Ashry:2013loa, Ashrythesis}, and more recently by Ref.\cite{Frank:2020odd} in exploring the Majorana nature of the right-handed neutrino through  $0\nu\beta\beta$ and possible leptogenesis.

The scalar sector of this model consists of one Higgs bidoublet $\Phi$ and two Higgs doublets (one each under $SU(2)_L$ and $SU(2)_R$) $\chi_{L,R}$. With multiple scalar fields present, the parameter space needs to be analysed carefully to establish the viable ranges values that can provide a  stable vacuum. In the conventional LRSM, the Higgs bidoublet is joined by two triplet scalar fields.  General techniques on the boundedness of the potential from below and copositivity condition \cite{Kannike:2012pe,Kannike:2016fmd,Sanchez-Vega:2018qje} on the quartic couplings were discussed for multi-scalar potentials in Refs.\cite{Chakrabortty:2013zja, Chakrabortty:2013mha}, which then apply these conditions to the specific case of LRSM. Arguing that these conditions are necessary but not sufficient, in a recent study the structure of the vacuum of LRSM was analysed to obtain the constrains in the parameter space, and their consequences in the mass spectrum fo the physical scalar of the model \cite{BhupalDev:2018xya}.  A more general approach to find the stable vacuum with the scalar sector of LRSM was performed in Ref.\cite{Chauhan:2019fji} employing the concept of copositivity and gauge orbit space \cite{Kim:1981xu, Kannike:2016fmd}. An alternative analysis of the vacuum stability of the left-right symmetric model is given in \cite{Kannike:2021fth}. Some other references relevant to vacuum structure of multiple scalar potential include \cite{Kim:1981xu,Branco:1985ng,Basecq:1985sx,Mohapatra:2014qva,Brahmachari:1994ts}.

In this work, we identify the desirable vacuum structure and its stability in terms of 2 scalar mass values duly constrained from data, and 8 quartic and trilinear scalar couplings. The structure of the paper is as follows. We start with the particle content description of ALRM in context of DLRM2 scenario in Section \ref{sec:model}. The scalar potential involving a bidoublet and doublet Higgs representations together with the minimisation conditions are discussed in Section \ref{sec:potential}. In Section \ref{sec:copositivity} we perform a general analytic study of the vacua looking for the sufficient and necessary conditions to have a connection between the electromagnetic charge-preserving and electromagnetic charge-breaking vacua, differentiating the contributions of the bidoublet, doublets and from the coupled parts of the scalar potential separately. In Section  \ref{sec:masses} we look at the implications on the parameters in the scalar potential coming from bounds on neutral and charged Higgs masses. We make some final observations connecting all the constraints in Section \ref{sec:comments} and conclude in Section \ref{sec:conclusion}.

\section{Description of the Alternative Left-Right Model}
\label{sec:model}
The gauge structure of the ALRM discussed in the previous has the symmetry group
\begin{center}
    $\mathcal{G}_{ALRM} \equiv SU(3)_C \otimes SU(2)_L \otimes SU(2)_{R^\prime} \otimes U(1)_{B-L} \otimes S$
\end{center}
with all except $S$ gauged. The prime on $SU(2)_{R^\prime} $ is to remind us that the right-handed fermion doublet in this case is different from that of the conventional LRSM, in the sense that the right-handed partner of the down-type quarks and the right-handed neutrinos are singlets under this. 
The particle content of the model is given in Table \ref{tab_ALRM}, where we have considered the $S$ charge within the DLRM3 \cite{Ma:2012gb} for specificity. Notice that the scalar sector of all the three versions remain the same, and therefore the main discussion in this work regarding the vacuum structure of the ALRM is application as it is to all the versions.
\begin{table}[htb!]
\vskip 5mm
\centering
{
\begin{tabular}{|c|c|c|c|c|c|}
 \hline \hline
Particles&{$SU(3)_C$} & $SU(2)_L$ & $SU(2)_{R'}$ & $U(1)_{B-L}$ & S
           \\[2mm]
\hline 
\multicolumn{6}{l}{Quarks}\\
\hline 
 $Q_{L} = 
\begin{pmatrix}
u_L \\
d_L
\end{pmatrix}$ & 3 & 2 & 1 & $\frac{1}{6}$ & 0
          \\[2mm]
$Q_{R} = 
\begin{pmatrix}
u_R \\
d'_R
\end{pmatrix}$ & 3 & 1 & $2$ & $\frac{1}{6}$ & $-\frac{1}{2}$
          \\[2mm]
$d'_L$ & 3 & 1 & 1 & $-\frac{1}{3}$ & -1
          \\[2mm]
$d_R$ & 3 & 1 & 1 & $-\frac{1}{3}$ & 0
          \\[2mm]
\hline
\multicolumn{6}{l}{Leptons}\\
\hline 
 $L_{L} = 
\begin{pmatrix}
\nu_L \\
e_L
\end{pmatrix}$ & 1 & 2 & 1 & $-\frac{1}{2}$ & 0
          \\[2mm]
 $L_{R} = 
\begin{pmatrix}
n_R \\
e_R
\end{pmatrix}$  & 1 & 1 & 2 & $-\frac{1}{2}$ & $+\frac{1}{2}$
          \\[2mm]
$n_L$ & 1 & 1 & 1 & 0 & 1
          \\[2mm]
$\nu_R$ & 1 & 1 & 1 & 0& 0
          \\[2mm]
\hline
\multicolumn{6}{l}{Scalars}\\
\hline 
 $\Phi = 
\begin{pmatrix}
\phi_1^0 & \phi_1^+\\
 \phi_2^- & \phi_2^0
\end{pmatrix}$ & 1 & 2 & $ 2^{\ast} $ & 0 & $-\frac{1}{2}$
          \\[2mm]
 $\chi_{L} = 
\begin{pmatrix}
\chi_L^+ \\
\chi_L^0
\end{pmatrix}$ & 1 & 2 & 1 & $\frac{1}{2}$ & 0
        \\[2mm]
 $\chi_{R} = 
\begin{pmatrix}
\chi_R^+ \\
\chi_R^0
\end{pmatrix}$ & 1 & 1 & 2 & $\frac{1}{2}$ & $\frac{1}{2}$
        \\[2mm]
        \hline 
\multicolumn{6}{l}{Gauge Bosons}\\
\hline 
 $G^\mu$ & 8 & 1 & 1 & 0 & 0 \\[2mm]
 $W^\mu_L$ & 1 &3 & 1 &0 & 0 \\[2mm]
 $W^\mu_R$ & 1 & 1 & 3 & 0 & 0 \\[2mm]
 $B^\mu$ & 1 & 1 & 1 & 0 & 0 \\[2mm]
  \hline\hline
\end{tabular}
}
\caption{Particle content of ALRM.}
\label{tab_ALRM}
\end{table}
The standard left-handed (LH) fermions form $SU(2)_L$ doublets, while the right-handed (RH) up-type quarks with down-type quarks $d_R^\prime$ form RH doublets. Similarly, RH charged leptons partner with exotic RH neutral fermions ($n_R$) to form $SU(2)_{R^\prime}$ doublets. In addition, there are the left-handed ($d_L'$) and right-handed ($d_R$) quarks and two non-coloured neutral fermions, $n_L$ and $\nu_R$, which are singlets under both $SU(2)_L$ and $SU(2)_{R'}$.  The Yukawa interactions and the symmetry breaking to the SM gauge group are such that the Dirac mass partner of $d_L$ is $d_R$, not $d'_R$, and that of $\nu_L$ is $\nu_R$, not $n_R$ \cite{Ma:2012gb}.  We shall see that along with the $SU(2)_R$ symmetry breaking the global symmetry $S$ is broken such that a combination of $S'=S+T_{3R}$ is unbroken. Further breaking of the electroweak symmetry preserves this as long as $\phi_1^0$ does not acquire vacuum expectation value. We shall see in the following that this can be arranged in a natural way.   The right-handed neutral fermion, $n_R$ is a dark matter candidate, as explored in Refs.\cite{Frank:2004vg, Frank:2019nid, Ma:2010us, Khalil:2009nb, Khalil:2010yt, Ma:2012gb}.  With soft-breaking of the lepton number, a Majorana mass term can be considered for $\nu_R$, and the consequent lepton asymmetry could lead to leptogenesis \cite{Frank:2020odd}.  In addition, allowing soft-breaking of $S'$ a Majorana mass term can be considered for $n_R$, thus contributing to possible leptogenesis \cite{Ma:2012gb}.  In this work, we shall not discuss  these dynamics any further. Rather, we shall focus our attention to the scalar sector and study the constrains emerging to establish a stable vacuum structure.

The scalar sector of the model consists of one bidoublet $\Phi$ Higgs field with the $SU(2)_L$ symmetry acting along the column and the $SU(2)_R$ along the row as represented in Table~\ref{tab_ALRM}.  In addition, there are  two Higgs fields transforming as doublets under $SU(2)_L$ and $SU(2)_{R^\prime}$, denoted by $\chi_{L}$ and $\chi_R$, respectively.  The VEV of $\chi_R$ breaks $SU(2)_R\otimes U(1)_{B-L}$ to $U(1)_Y$, and the  subsequent breaking of the electroweak symmetry is driven  by the VEVs of $\Phi$ and $\chi_L$. In presence of extra the $S$ symmetry, quark doublets can interact with $\tilde{\Phi}$ and lepton doublets with $\Phi$ only. With this differentiation one can avoid the unwanted mixing between $W_L-W_R$ gauge boson mixing as well as $d, d^\prime$ and $n, \nu$ mixing in the model. Collider phenomenology of the model studying the 
Higgs production and decays are discussed in \cite{Ashry:2013loa,Ashrythesis}, and collider searches of heavy gauge bosons and constrains derived from these on the model parameters are elaborately discussed in \cite{Frank:2019nid}.  In the present work we concentrate on the scalar potential to explore the vacuum structure of the ALRMs.

\section{The Scalar Potential of ALRM and minimization}
\label{sec:potential}
The most general scalar potential respecting the gauge symmetries and global $U(1)_S$ in ALRM can be written as
\begin{align}
\mathcal{V}_H & = - \mu_1^2 \text{Tr} \left[\Phi^{\dagger} \Phi \right] - \mu_2^2 \left(\chi_L^{\dagger} \chi_L +  \chi_R^{\dagger} \chi_R \right)
+ \lambda_1 \left( \text{Tr} \left[\Phi^{\dagger} \Phi \right] \right)^2
+  \lambda_2 \text{Tr} \left[\tilde{\Phi}^{\dagger}  \Phi \right] \text{Tr} \left[\Phi^{\dagger} \tilde{\Phi} \right]
  \nonumber \\
& + \rho_1 \left[ \left(\chi_L^{\dagger} \chi_L \right)^2 +  \left(\chi_R^{\dagger} \chi_R \right)^2 \right] 
+ 2 \rho_2 \left(\chi_L^{\dagger} \chi_L\right) \left(\chi_R^{\dagger} \chi_R \right) 
+2 \alpha_1 \text{Tr} \left[\Phi^{\dagger} \Phi \right] \left(\chi_L^{\dagger} \chi_L +  \chi_R^{\dagger} \chi_R \right)
\nonumber \\
& +2 \alpha_2 \left[ \chi_L^{\dagger} \Phi \Phi^\dagger \chi_L + \chi_R^{\dagger} \Phi^\dagger \Phi \chi_R \right]
+2 \alpha_3 \left[ \chi_L^{\dagger} \tilde{\Phi} \tilde{\Phi}^\dagger \chi_L + \chi_R^{\dagger} \tilde{\Phi}^\dagger \tilde{\Phi} \chi_R \right]
+ \mu_3 \left[\chi_L^\dagger \Phi \chi_R + \chi_R^\dagger \Phi^\dagger \chi_L \right] \, 
\label{eq:pot}
 \end{align}
where we take all the couplings to be real as we restrict our analysis to the  CP-conserving  case. $SU(3)_C$ singlets, the  Higgs fields  transformation under other gauge groups as
\begin{eqnarray}
SU(2)_L \otimes SU(2)_R &:& \Phi \rightarrow U_L \Phi U_R^{\dagger},~~ \chi_L \rightarrow U_L \chi_L, ~~ \chi_R \rightarrow U_R \chi_R \nonumber \\
U(1)_{B-L}  &:&  \Phi \rightarrow  \Phi ,~~ \chi_L \rightarrow e^{i{\frac{1}{2}}\theta_{B-L}} \chi_L, ~~ \chi_R \rightarrow e^{i\frac{1}{2}\theta_{B-L}} \chi_R
\end{eqnarray}
where $U_{L,R} $ are ($ 2 \times 2$) transformation matrices corresponding to  $SU(2)_L$ and $SU(2)_R$, respectively, and $e^{i\frac{1}{2}\theta_{B-L}}$ is the phase factor characterising transformations under $ U(1)_{B-L}$. We note here that the dual field $\tilde{\Phi} \equiv \sigma_2 \Phi^{\ast} \sigma_2$ transforms in the same way as $\Phi$ under the gauge symmetry, but carry opposite $S$ charge.  Similar relations hold for  the duals of the doublets,  
 $\tilde{\chi}_{L,R} = i\sigma_2 \chi_{L,R}^\ast$.

As mentioned in the previous section, VEV of $\chi_R$ breaks $SU(2)_R\otimes U(1)_{B-L}$ to $U(1)_Y$ and further electroweak symmetry is broken by the VEVs of the bidoublet $\Phi$ and the left-handed doublet $\chi_L$. 
In general, the electromagnetic charge conserving VEVs of the scalar fields should have the form 
\begin{equation}
\langle \Phi \rangle = \frac{1}{\sqrt{2}} 
\begin{pmatrix}
v_1 e^{i \theta_1}& 0 \\
0 & v_2 e^{i \theta_2}
\end{pmatrix}, 
~~\langle \chi_L \rangle = \frac{1}{\sqrt{2}} 
\begin{pmatrix}
0 \\
v_L e^{i \theta_L}
\end{pmatrix}, 
~~\langle \chi_R \rangle = \frac{1}{\sqrt{2}} 
\begin{pmatrix}
0 \\
v_R
\end{pmatrix}\, ,
\label{eq:VEVs}
\end{equation}
where $\theta_1,~\theta_2, ~\theta_L$ are the phases in the scalar sector which cannot be rotated away by the field redefinitions.  Requiring that the potential has a minimum for the VEVs given by Eq. \ref{eq:VEVs},  the minimization conditions are
\begin{equation}
    \frac{\partial \mathcal{V}_H}{\partial v_1} =  \frac{\partial \mathcal{V}_H}{\partial v_2} =\frac{\partial \mathcal{V}_H}{\partial \theta_1}=  \frac{\partial \mathcal{V}_H}{\partial \theta_2} =  \frac{\partial \mathcal{V}_H}{\partial \theta_L} =  \frac{\partial \mathcal{V}_H}{\partial v_L} = \frac{\partial \mathcal{V}_H}{\partial v_R} = 0\, ,
\end{equation}
where the potential is taken at the vacuum configuration of the fields in  Eq. \ref{eq:VEVs}. Taking the derivatives with respect to $v_1$, one of the minimization conditions read
     
\begin{align}
    \frac{\partial \mathcal{V}_H}{\partial v_1}  &= -\mu_1^2 v_1 + \lambda_1 (v_1^2 + v_2^2) v_1 + 2\lambda_2 v_2^2 v_1 + (v_L^2 + v_R^2)(\alpha_1 + \alpha_3) v_1 = 0 \, 
    \label{chi1} 
    \end{align}
    Further, Eq. \ref{chi1} allows us to set $v_1 = \langle \phi_1^0 \rangle = 0$,  resulting in vanishing of unwanted $W_L-W_R$ mixing in ALRM and making the angle $\theta_1$ irrelevant. This also leads to the  decoupling of CP-even and CP-odd components of  $\phi_1^0$ without mixing with other Higgs fields. The other minimization conditions are, setting $v_1=0$,
 \begin{align}
    \frac{\partial \mathcal{V}_H}{\partial v_2}  &= -\mu_1^2 v_2 + \lambda_1  v_2^3   + + (v_L^2 + v_R^2)(\alpha_1 + \alpha_2) v_2 + \frac{\mu_3}{\sqrt{2}}v_L v_R \text{cos}(\theta_2 -\theta_L) = 0 
    \label{chi2}\\
    \frac{\partial \mathcal{V}_H}{\partial v_L}  &=  -\mu_2^2 v_L + \rho_1 v_L^3 + \rho_2 v_L v_R^2 + \alpha_1  v_2^2v_L + \alpha_2 v_L v_2^2  + \frac{\mu_3}{\sqrt{2}} v_R v_2 \text{cos}(\theta_2 -\theta_L) = 0
     \label{vL}\\
    \frac{\partial \mathcal{V}_H}{\partial v_R} &=  -\mu_2^2 v_R + \rho_1 v_R^3 + \rho_2 v_L^2 v_R + \alpha_1  v_2^2v_R + \alpha_2 v_R v_2^2  + \frac{\mu_3}{\sqrt{2}} v_L v_2 \text{cos}(\theta_2 -\theta_L) = 0 
    \label{vR}\\
      \frac{\partial \mathcal{V}_H}{\partial \theta_2}  &=  \frac{\partial \mathcal{V}_H}{\partial \theta_L} = \frac{\mu_3}{\sqrt{2}} v_2 v_L v_R \text{sin}(\theta_2 -\theta_L) = 0 
  \label{eq:theta}
\end{align}
Eqs. \ref{eq:theta}  imply $\theta_2 = \theta_L$ as $\mu_3, v_2, v_{L,R} \neq 0$, effectively removing one of the remaining phases, and leaving a single phase arising from the vacuum structure of ALRM. This is in contrast to the conventional LRSM having multiple phases in the vacuum configuration.
Setting  $\theta_2=\theta_L$ in Eqs. \ref{vL} and \ref{vR}, we have
\begin{align}
 -\mu_2^2 + \rho_1 v_L^2 + \rho_2 v_R^2 + \alpha_1  v_2^2 + \alpha_2  v_2^2 &= - \frac{\mu_3}{\sqrt{2}} \frac{v_R}{v_L} v_2 
     \label{vL2}\\
 -\mu_2^2 + \rho_1 v_R^2 + \rho_2 v_L^2  + \alpha_1  v_2^2+ \alpha_2 v_2^2  &= - \frac{\mu_3}{\sqrt{2}} \frac{v_L}{v_R} v_2 
    \label{vR2}
\end{align}
Subtracting Eq. \ref{vR2} from \ref{vL2} gives
\begin{equation}
    (v_R^2 - v_L^2)\left(\frac{\mu_3}{\sqrt{2}} \frac{v_2}{v_L v_R} -  (\rho_1 -\rho_2)\right) = 0 
\end{equation}
For  $v_L \neq v_R$,  (required to break left-right symmetry), we can express the trilinear coupling in terms fo the VEVS and the quartic couplings involving only the doublet scalar fields as
\begin{equation}
\boxed{
\mu_3= \frac{\sqrt{2}~v_L v_R (\rho_1 -\rho_2)}{v_2}}
    \label{seesaw}
\end{equation}
The other relations found from the minimization conditions are
\begin{equation}
\boxed{
    v_L^2 + v_R^2 = \frac{\mu_2^2 - \left( \alpha_1 + \alpha_2 \right)v_2^2}{\rho_1}}\, ,
\end{equation}
and
\begin{equation}
\boxed{
    v_2^2 = \frac{2 \left[\left( \rho_1 \mu_1^2 - (\alpha_1 + \alpha_2)\mu_2^2 \right)  (\rho_2 - \rho_1) \right] + \rho_1 \mu_3^2 }{2 \left[\lambda_1 \rho_1 - (\alpha_1 - \alpha_2)^2 \right] (\rho_2 - \rho_1)}}\, .
\end{equation}

\subsection{Generalised Vacuum Structure}

While a realistic vacuum structure is required to respect the $U(1)_{em}$ symmetry, and consequently, require to have zero VEV for all electrically charged fields, it may be informative to explore the vacuum structure by relaxing this criteria. As pointed out in Ref.\cite{BhupalDev:2018xya} in the case of conventional LRSM with triplet scalar fields along with the bidoublet, some of these electromagnetic charge-breaking vacua could be connected to the electromagnetic charge-preserving vacuum through a gauge transformation.  To demonstrate this in the case of ALRM, we first use the gauge symmetries to remove some of the degrees of freedom and restrict the scalar fields to 
\begin{equation}
\Phi  =
\begin{pmatrix}
\phi_1 e^{-i\theta_1} & 0 \\
0 & \phi_2 e^{-i\theta_2}
\end{pmatrix}, 
~~ \chi_L =  
\begin{pmatrix}
b_1 \\
a_1 e^{i \theta_L}
\end{pmatrix}, 
~~ \chi_R  = 
\begin{pmatrix}
b_2  \\
a_2 
\end{pmatrix}.
\label{bad_vev_fields}
\end{equation}
where $\phi_{1,2},~a_{1,2}$ and $b_{1,2}$ are real quantities.
In this way, the vev of the bidoublet always respects the $U(1)_{em}$ symmetry, whereas that is not the case with the doublets. 
Let us seek global gauge transformations $U_L$, $U_R$ which can transform the above vev's into favorable ones.
We let $\langle \phi_{1}\rangle=\langle \phi_{2}\rangle \ne 0$, but seek transformations that leave $\langle \Phi\rangle$ invariant. Such are then required to be
\begin{center}
    $U_L = U_R = \frac{1}{\sqrt{2}} 
    \begin{pmatrix}
    1 & -1 \\
    1 & 1
    \end{pmatrix} \, .
    $
    \end{center}
We next seek their effect on the doublets with $\langle a_{1,2}\rangle \ne 0$ and the $U(1)_{em}$ breaking $\langle b_{1,2}\rangle \ne 0$. The result can be clubbed into two specific cases \\[5mm]       
    \textbf{Case 1 :} with $\langle a_{1,2}\rangle=\langle b_{1,2}\rangle=\frac{v_{L,R}}{\sqrt{2}}$  and $\langle \theta_L\rangle=0$, the doublets transform under the above $SU(2)_L\otimes SU(2)_R$ transformation as 
 \begin{eqnarray}
\left\{\langle \chi_{L,R} \rangle =\frac{v_{L,R}}{\sqrt{2}}\begin{pmatrix}
    1 \\
    1
    \end{pmatrix} \right\}~~~~\longrightarrow~~~~
\begin{pmatrix}
    0 \\
    v_{L,R}
    \end{pmatrix},
    \end{eqnarray}
   connecting the $U(1)_{em}$ violating vacuum to the charge-preserving vacuum through a gauge transformation. On the other hand, \\[5mm]
       \textbf{Case 2 :} with $\langle a_{1,2}\rangle=-\langle b_{1,2}\rangle=-\frac{v_{L,R}}{\sqrt{2}}$  and $\theta_L=0$, the doublets transform as 
 \begin{eqnarray}
\left\{\langle \chi_{L,R} \rangle =\frac{v_{L,R}}{\sqrt{2}}\begin{pmatrix}
    1 \\
    -1
    \end{pmatrix} \right\}~~~~\longrightarrow~~~~
\begin{pmatrix}
        v_{L,R}\\
        0
    \end{pmatrix},
    \end{eqnarray}
which does not yield the required desirable vacuum structure.

By inspection then we see that the "good" charge-preserving vacua can be characterised by demanding that there exist $\tilde{U}_L$, $\tilde{U}_R$, such that
\begin{equation}
\sigma^1 \tilde{U}_L\langle \chi_{L} \rangle = \langle \chi_{L} \rangle; \qquad \sigma^1\tilde{U}_R \langle \chi_{R} \rangle = \langle \chi_{R} \rangle
\end{equation}
where the Pauli matrix belongs to the relevant gauge group. We expect the criterion to continues to work when the simplifying condition $\langle \phi_{1}\rangle=\langle \phi_{2}\rangle$ is relaxed so long as the vacua sought are in a neighbourhood of the simpler case. This can be particularly useful in numerical calculations. However in the following analysis since we use VEV's belonging to $\mathbb{R}^+$, this criterion will be automatically incorporated.
     
\subsection{Analysis of the vacuum structure}
\label{sec:numerical}

\begin{figure}[h!]
\centering
\includegraphics[scale=0.5]{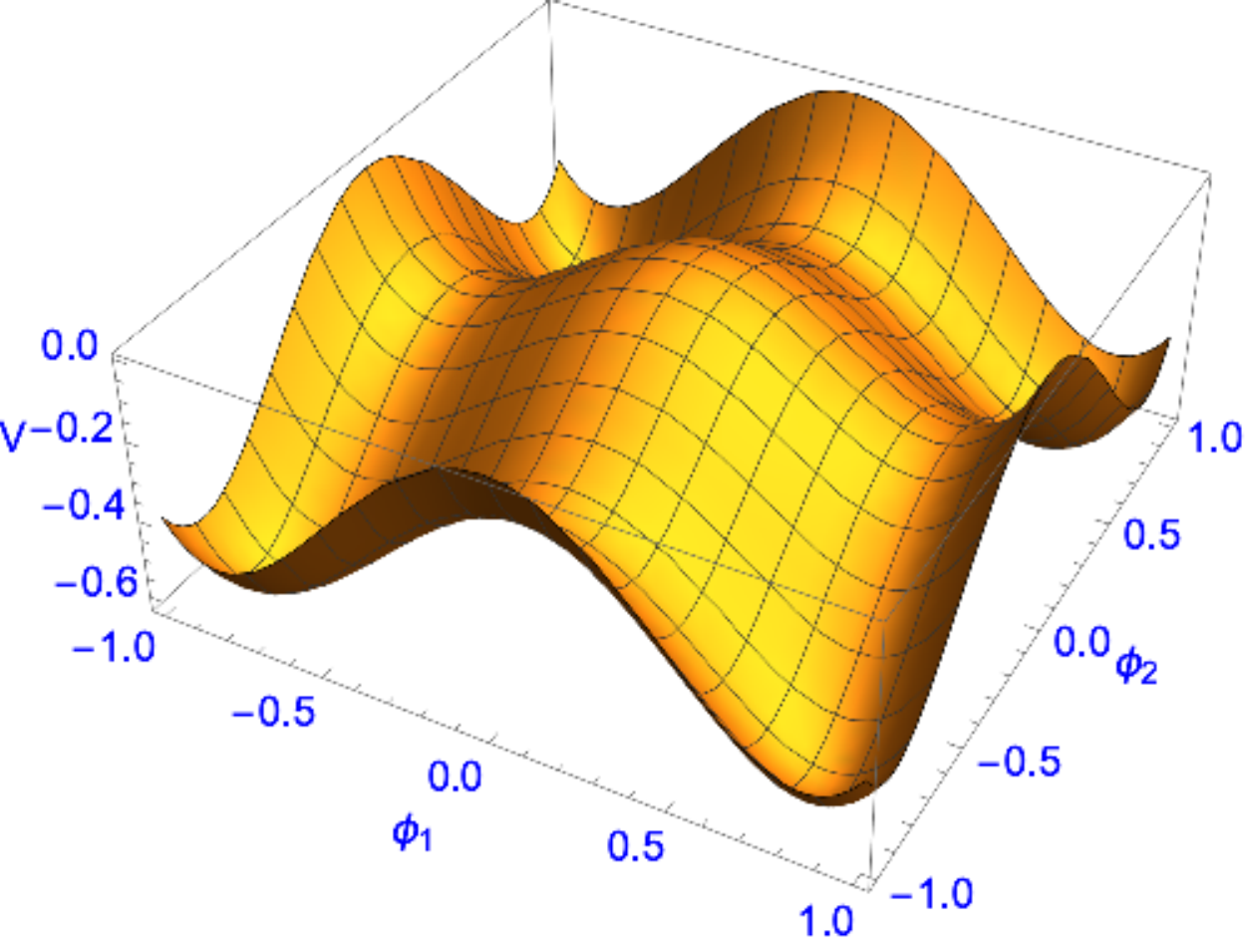} \qquad
\includegraphics[scale=0.4]{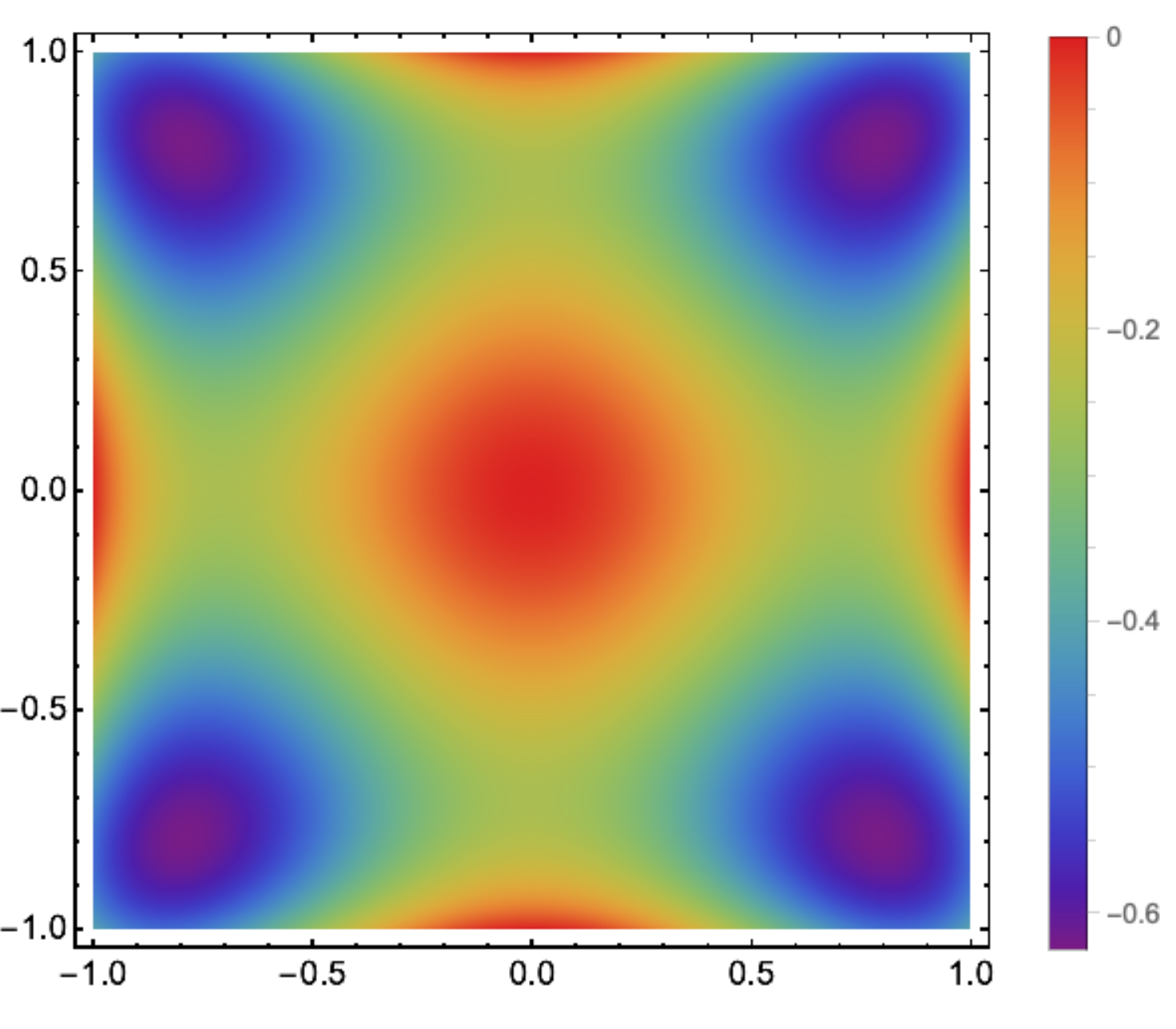}
\caption{Left Panel : The representation of the potential configuration for $\lambda_{12} < 0$. Right panel : the corresponding heat map for such configuration.}
\label{pot4}
\end{figure}

In this section, we follow our analytical investigations by a more detailed numerical analysis of the vacuum structure of the ALRM model and the implications on the parameter space. For the ease of analysis we shall consider a simplified case setting all $\alpha_i$ and $\mu_3$ coefficients to zero  to decouple the bidoublet from the doublets. This also allows a direct comparison with the LRSM case discussed in \cite{BhupalDev:2018xya}. With this, the scalar potential given in Eq. \ref{eq:pot} could be written as
\begin{equation}
\mathcal{V}_H |_{\substack{\alpha_i=0,\\\mu_3=0} }= \mathcal{V}_\Phi + \mathcal{V}_\chi
\label{eq:v_2pot}
\end{equation}
where $\mathcal{V}_\Phi$ and $\mathcal{V}_\chi$ contain only $\Phi$ and $\chi_{L/R}$ respectively. 
Plugging in the redefined fields as in Eq. \ref{bad_vev_fields} we have 
\begin{align}
    \mathcal{V}_\Phi (\phi_1, \phi_2) & = -\mu_1^2 (\phi_1^2 + \phi_2^2) + \lambda_1 (\phi_1^2 + \phi_2^2)^2 + 4\lambda_2 \phi_1^2 \phi_2^2
    \label{Vphi}
\end{align}
and
\begin{align}
    \mathcal{V}_\chi (a_1,a_2,b_1,b_2)  = &
    -\mu_2^2 \left[a_1^2+a_2^2+b_1^2+b_2^2\right] +\rho_1 \left[\left(a_1^2+b_1^2\right)^2+\left(a_2^2+b_2^2\right)^2\right] \nonumber\\
    &+2\rho_2 \left(a_1^2+b_1^2\right)\left(a_2^2+b_2^2\right)
\end{align}
Denoting $a_1^2+b_1^2\equiv \delta_1^2$ and $a_2^2+b_2^2\equiv \delta_2^2$
\begin{align}
    \mathcal{V}_\chi (\delta_1,\delta_2) & = 
    -\mu_2^2 \left(\delta_1^2+\delta_2^2\right)
    +\rho_1 \left(\delta_1^2+\delta_2^2\right)^2
    +2(\rho_2 - \rho_1)~\delta_1^2 \delta_2^2
\end{align}
One may notice that $\mathcal{V}_{\Phi}$ and $\mathcal{V}_{\chi}$ have same form, and with the replacement of $\mu_1^2 \rightarrow \mu_2^2, \lambda_1 \rightarrow \rho_1, 4\lambda_2 \rightarrow 2(\rho_2 - \rho_1)$ in $\mathcal{V}_{\Phi}$  one gets $\mathcal{V}_{\chi}$.
Further,  the potential $\mathcal{V}_{\Phi}$ obeys dihedral $\mathcal{D}_4$ symmetry 
\begin{equation}
\mathcal{D}_4 : (\phi_1 , \phi_2)^T \rightarrow \mathcal{R}(\phi_1 , \phi_2)^T 
\end{equation}
with 
\begin{equation}
\mathcal{R} = \begin{pmatrix}
0 & \pm 1\\
\pm 1 & 0
\end{pmatrix}, ~~{\rm or}~~ 
\begin{pmatrix}
 \pm 1 & 0 \\
0 & \pm 1
\end{pmatrix}\,.
\end{equation}
Conditions for $\mathcal{V}_{\Phi}$ to be bounded from below reads as $
    \lambda_1 >0,~~\lambda_1 > - \lambda_2$, as explained in Section \ref{subsec:bidoublet}. 
From the minimization conditions of $\mathcal{V}_\Phi$ we have 
\begin{equation}
 \left.  \frac{\partial \mathcal{V}_\Phi}{\partial \phi_1}\right|_{\phi_i=\langle \phi_i\rangle}= \phi_1 \left[ - \mu_1^2 + 2\lambda_1 \left(\phi_1^2 +\phi_2^2 \right) + 4\lambda_2 \phi_2^2 \right] = 0
\end{equation}
\begin{equation}
\left.  \frac{\partial \mathcal{V}_\Phi}{\partial \phi_2}\right|_{\phi_i=\langle \phi_i\rangle}= \phi_2 \left[ - \mu_1^2 + 2\lambda_1 \left(\phi_1^2 +\phi_2^2 \right) + 4\lambda_2 \phi_1^2 \right] = 0
\end{equation}
denoting the VEV by the same notation as that of the field, with possible solutions as 
\begin{enumerate}
\item  $\phi_1^2 =\phi_2^2 =\frac{\mu_1^2}{4(\lambda_1 + \lambda_2)}$ 
 \item $\phi_1^2  =\frac{\mu_1^2}{2\lambda_1},~~\phi_2=0$ 
  \item $\phi_1=0,~~\phi_2^2  =\frac{\mu_1^2}{2\lambda_1}$ 
\item $\phi_1=  \phi_2=0$ 
\end{enumerate}
Rather than working with $\lambda_1$ and $\lambda_2$ individually we will use the combination $\lambda_{12} \equiv \lambda_1 + 2\lambda_2$ to analyse the vacuum structure of the potential.
We distinguish three different cases : (i) $\lambda_{12 }> 0$, (ii) $\lambda_{12 }  = 0$ and (iii) $\lambda_{12 } < 0$.
\begin{enumerate}
\item  \underline{$\lambda_{12} > 0$} :  We take $\lambda_1 = 1.0$ throughout  whereas $\lambda_2$ can take several benchmark values while ensuring $\lambda_{12}  > 0$ (see the copositivity criteria later in Eq. \ref{eq:copositivity_lambda}). We distinguish three possible scenarios:
\begin{itemize}
\item (a) $\lambda_2 = 0 :$ The potential will have infinitely-many degenerate minima located circularly around $\lbrace \phi_1, \phi_2 \rbrace = (0,0)$, as shown in Fig. \ref{pot1} left, corresponding to the density plot for the potential shown  in Fig. \ref{pot2}, left diagram. 
\item (b) \underline{$\lambda_2 > 0$}: For such benchmark value  the corresponding scalar potential has 4 global minima $\lbrace \phi_1, \phi_2 \rbrace \simeq (0,\pm 0.7)$ and $(\pm 0.7, 0)$ with an overlapping region which can be considered as local minimum of the potential. The central region of potential in $\phi_1-\phi_2$ plane is flat compared to $\lambda_2 = 0$ case (Fig. \ref{pot1} middle). The heat map for such configuration is shown in Fig. \ref{pot2}, middle diagram. 
\item (c)  \underline{$-0.5 < \lambda_2 < 0 $}: Considering $\lambda_2$ value lies between $(-0.5,0.0)$, there can exist 4 distinct minima at around $\lbrace \phi_1, \phi_2 \rbrace = (\pm 0.7, \pm 0.7)$ (Fig. \ref{pot1} right). We show the corresponding density plot  in right diagram in Fig. \ref{pot3}. 
\end{itemize}
\begin{figure}[ht]
\centering
\includegraphics[scale=0.35]{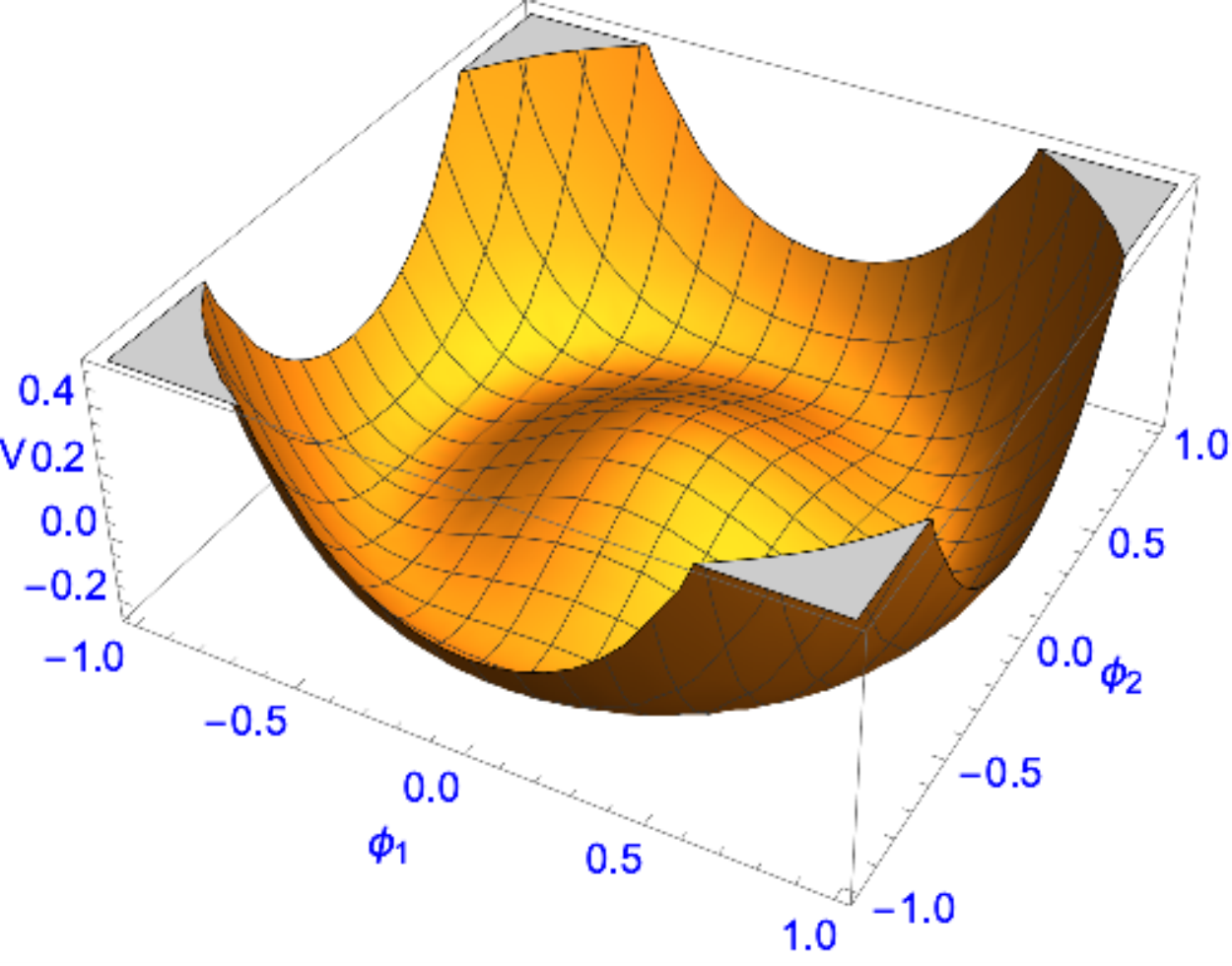}\qquad
\includegraphics[scale=0.35]{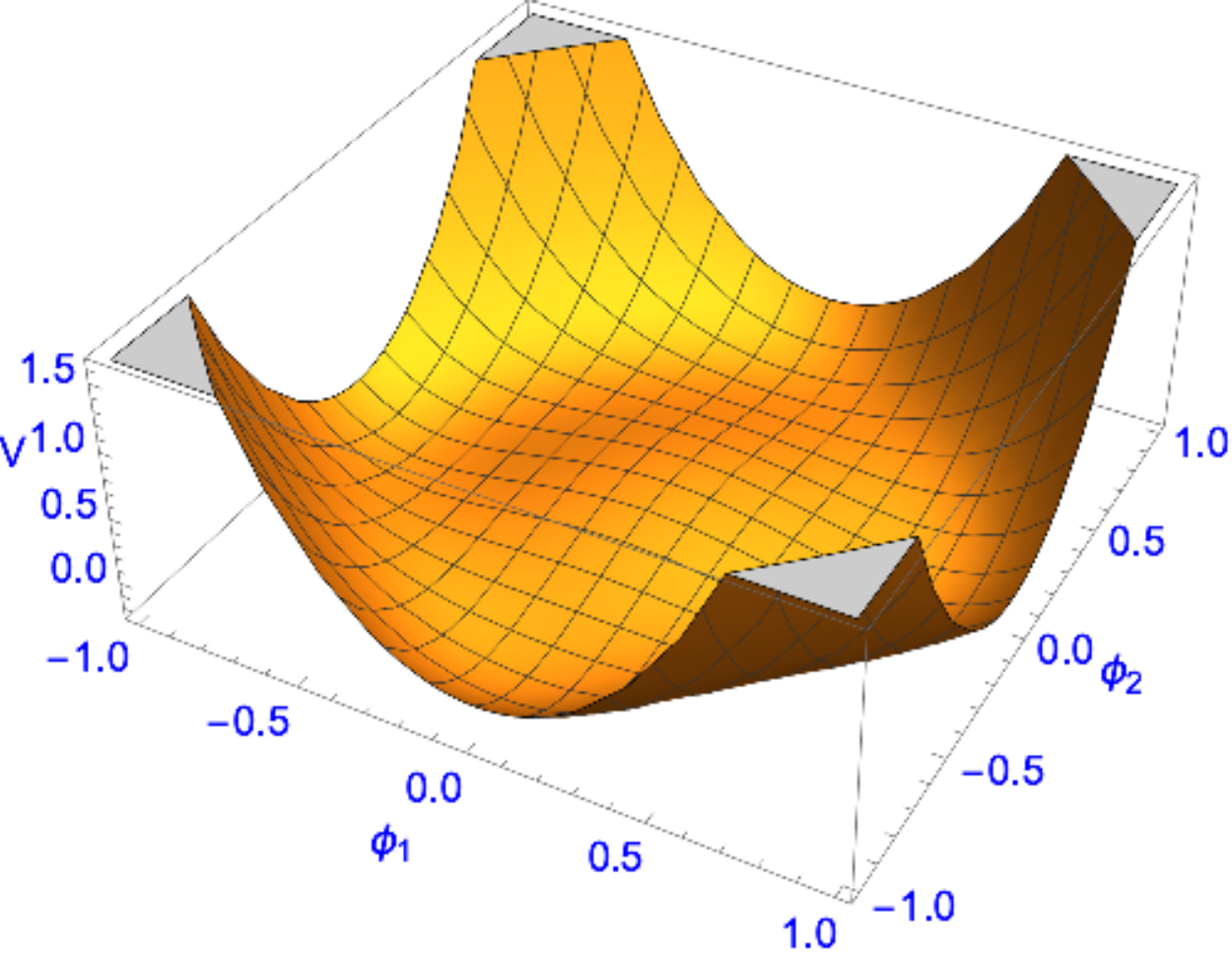}\qquad
\includegraphics[scale=0.35]{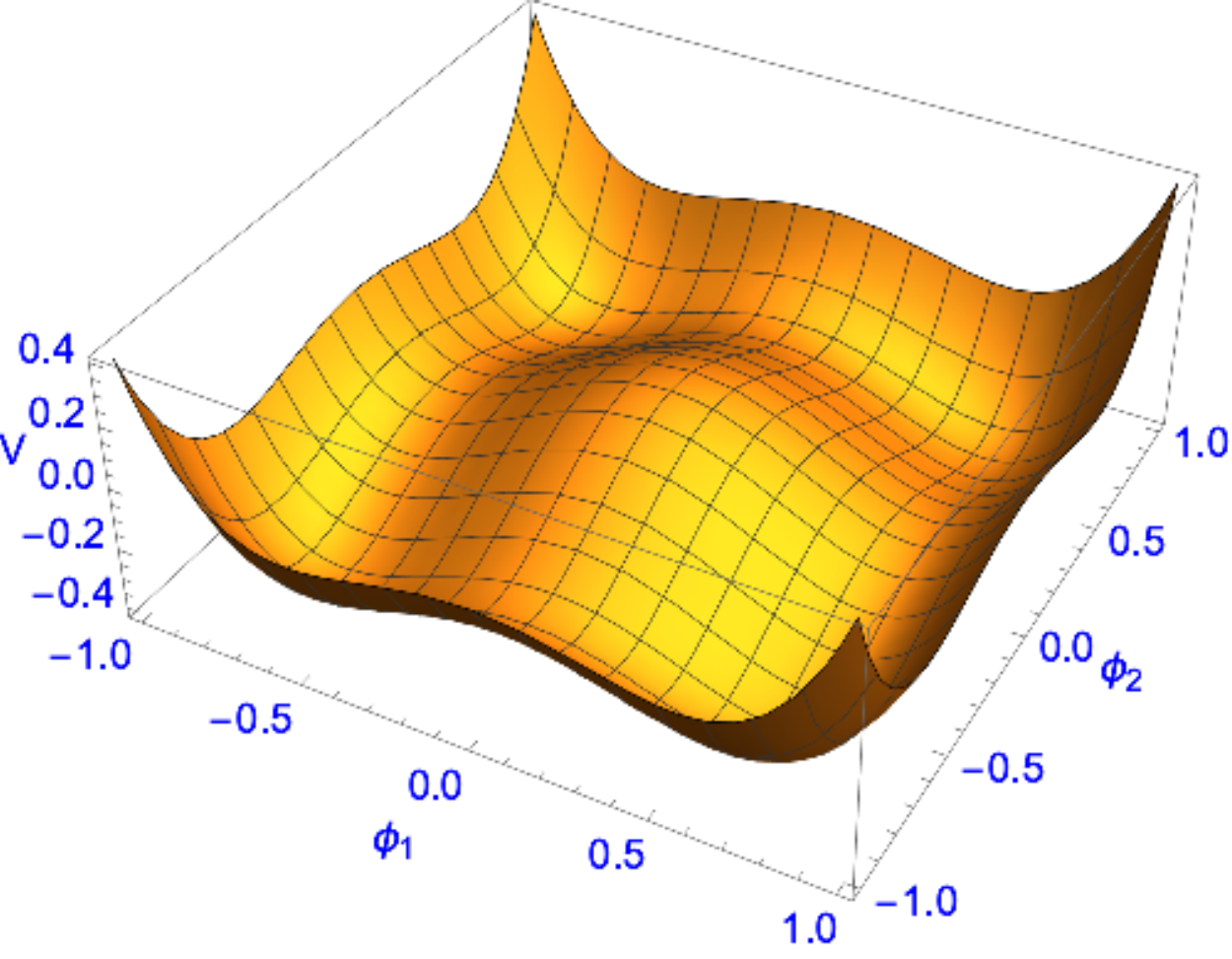}
\caption{Potential configurations in the $\phi_1-\phi_2$ plane with three possible choices of $\lambda_2$ while keeping $\lambda_{12} > 0$. Left : $\lambda_2 = 0$; middle : $\lambda_2 > 0$ and right : $-0.5 < \lambda_2 < 0$.}
\label{pot1}
\end{figure}

\begin{figure}[h!]
\centering
\includegraphics[scale=0.3]{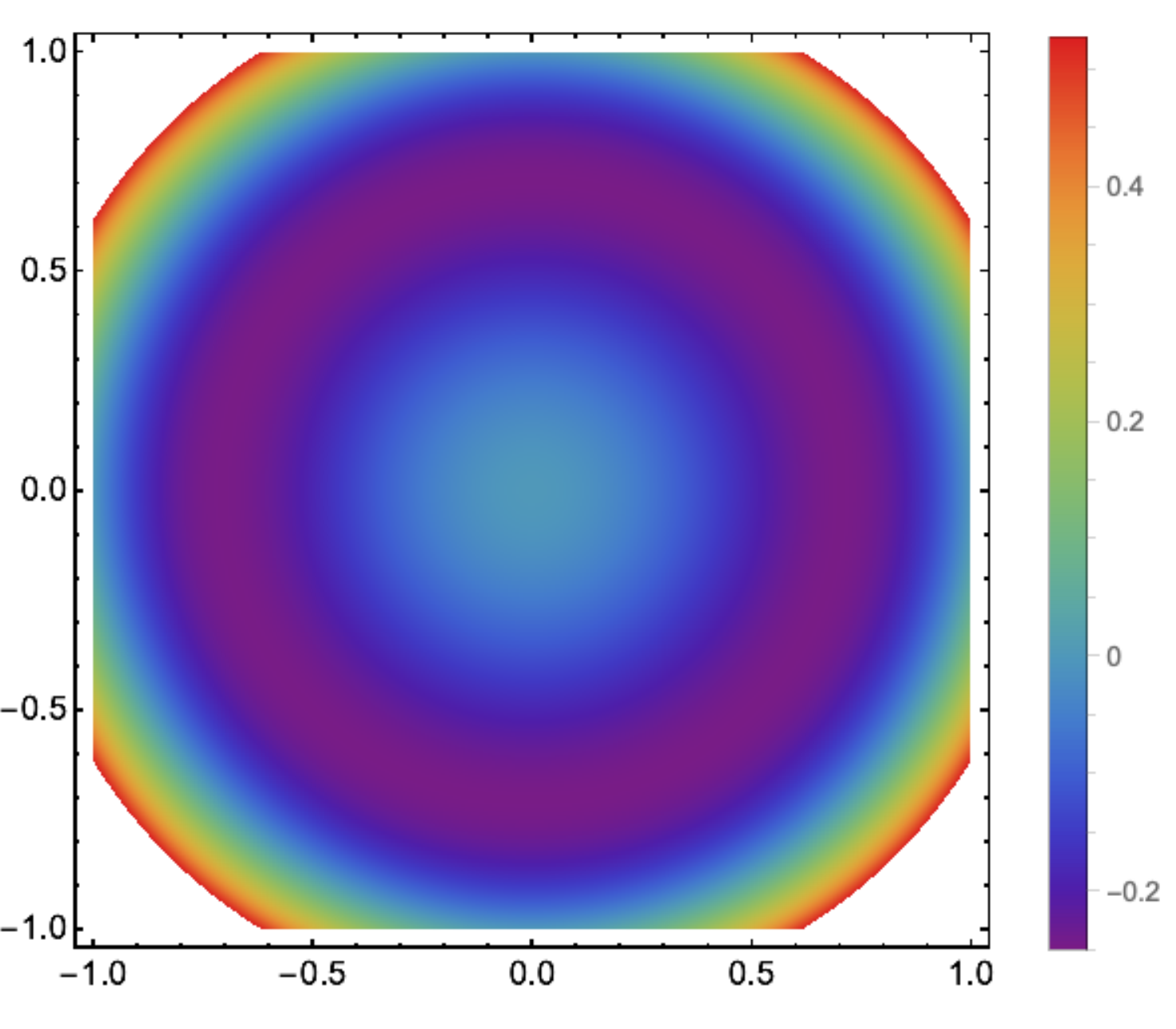}\qquad
\includegraphics[scale=0.3]{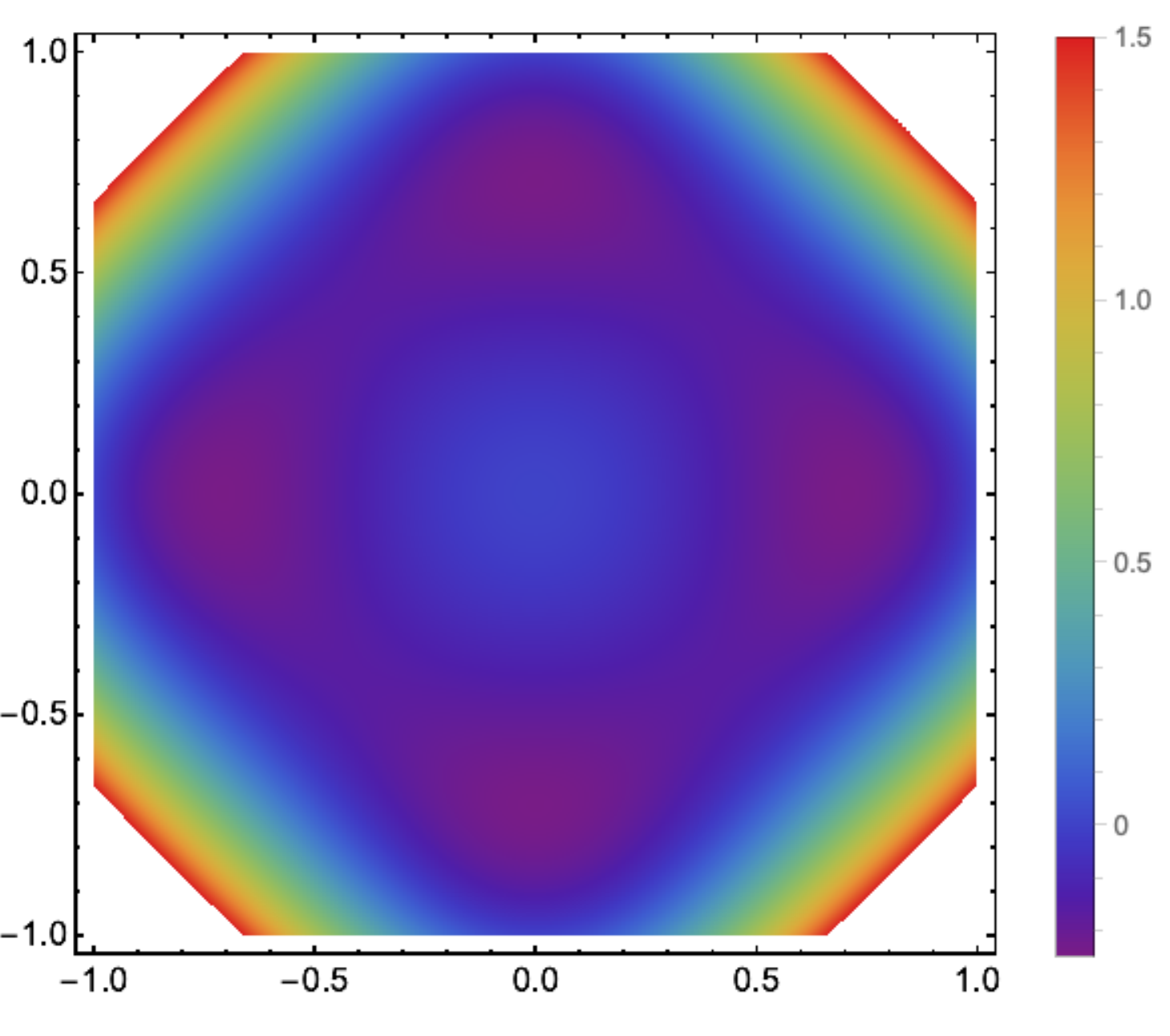}\qquad
\includegraphics[scale=0.3]{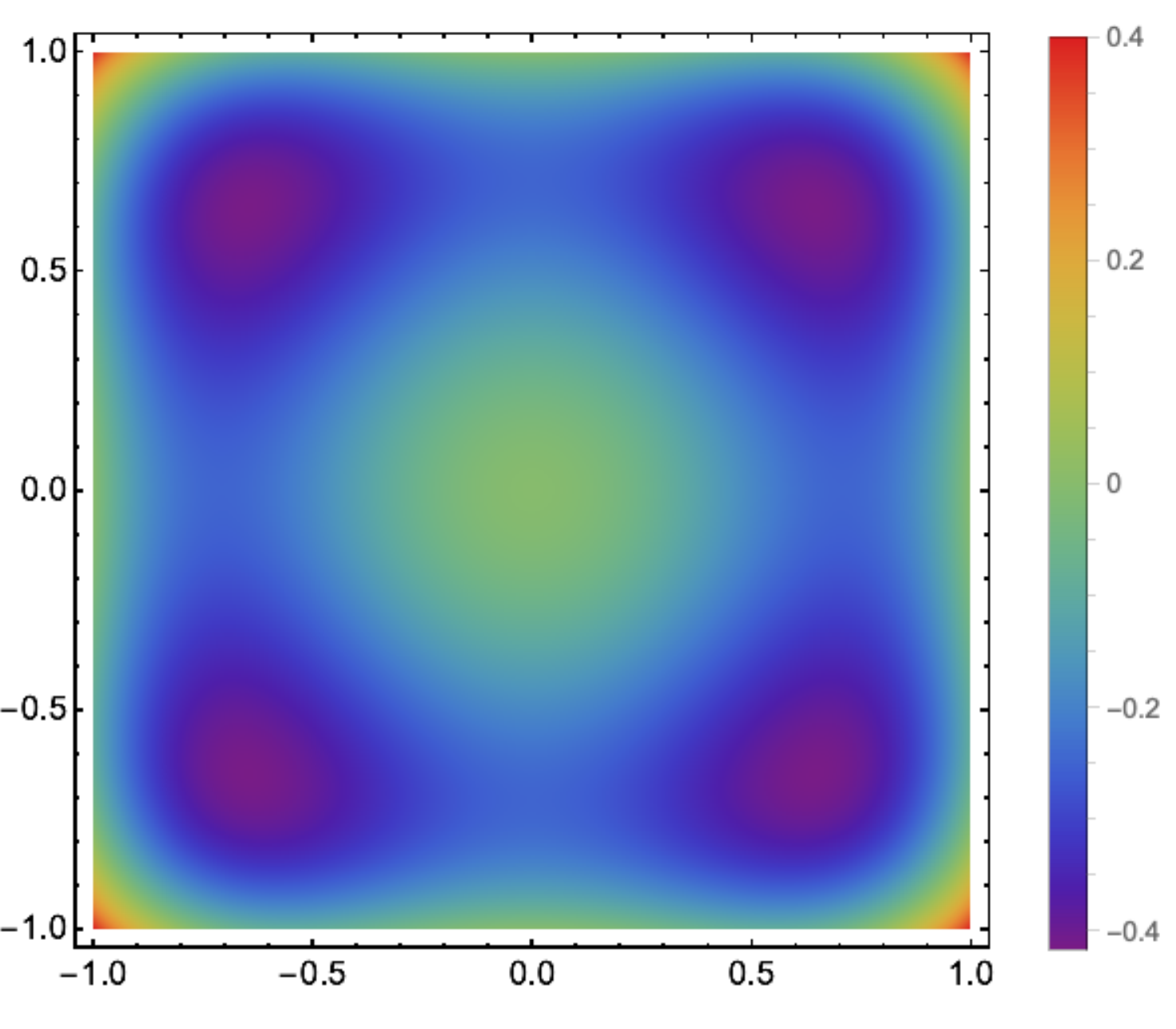}
\caption{Density plots for the potential configuration in the $\phi_1-\phi_2$ plane  for $\lambda_{12} > 0$. Left : $\lambda_2 = 0$; middle : $\lambda_2 > 0$ and right : $-0.5 < \lambda_2 < 0$.}
\label{pot2}
\end{figure}

\item \underline{$\lambda_{12} = 0$} : This condition is only satisfied for $\lambda_1 =1, \lambda_2 = -0.5$. In this scenario the potential has 4 distinct minima at around $\lbrace \phi_1, \phi_2 \rbrace = (\pm 0.7, \pm 0.7)$ (Fig. \ref{pot3} left panel), but the height between central bulge and corresponding minima has been increased with respect to the $\lambda_{12} > 0, -0.5< \lambda_2 <0$ case. We show, in right panel of Fig. \ref{pot3},  the density plot for  this case. 
\begin{figure}[h!]
\centering
\includegraphics[scale=0.5]{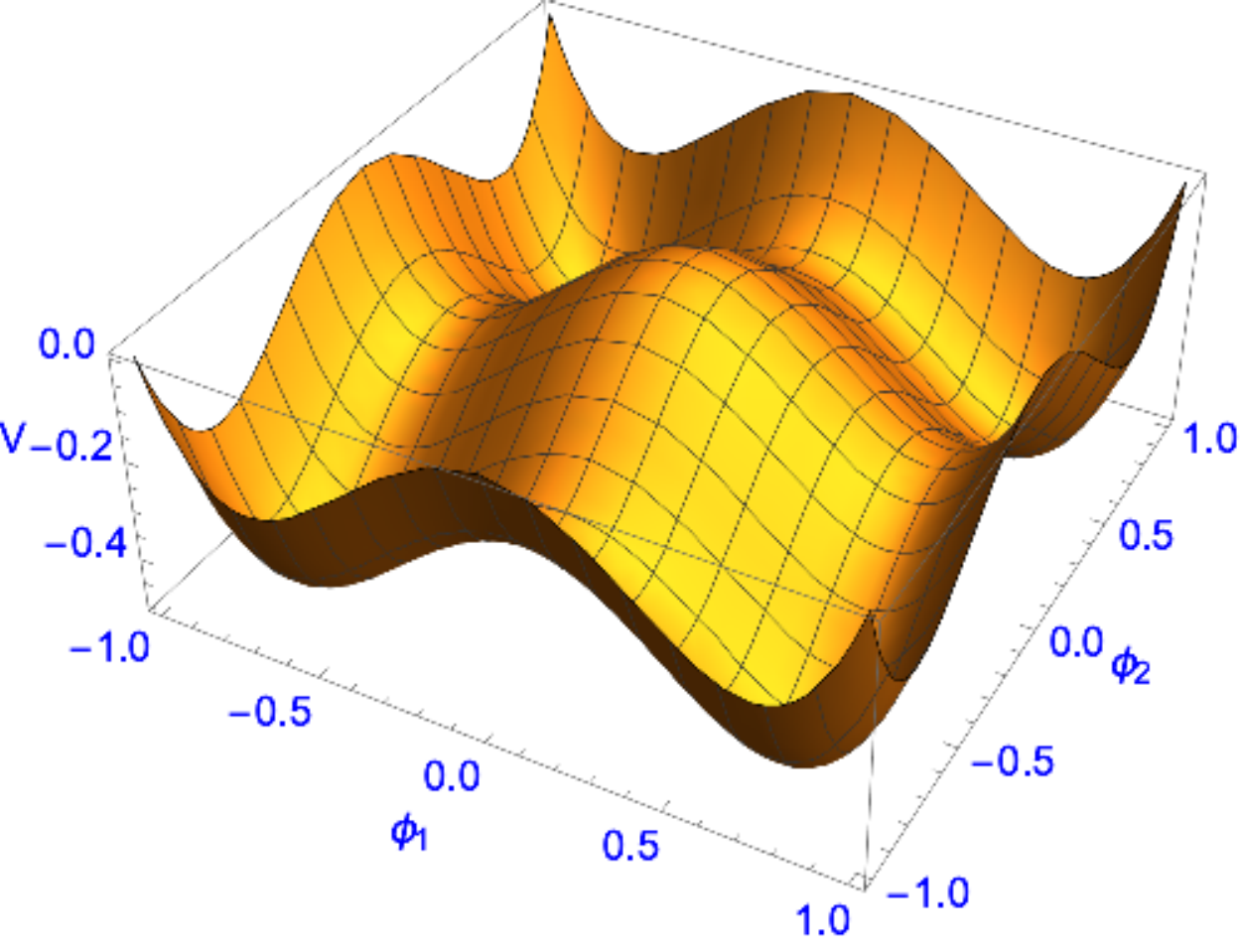} \qquad
\includegraphics[scale=0.4]{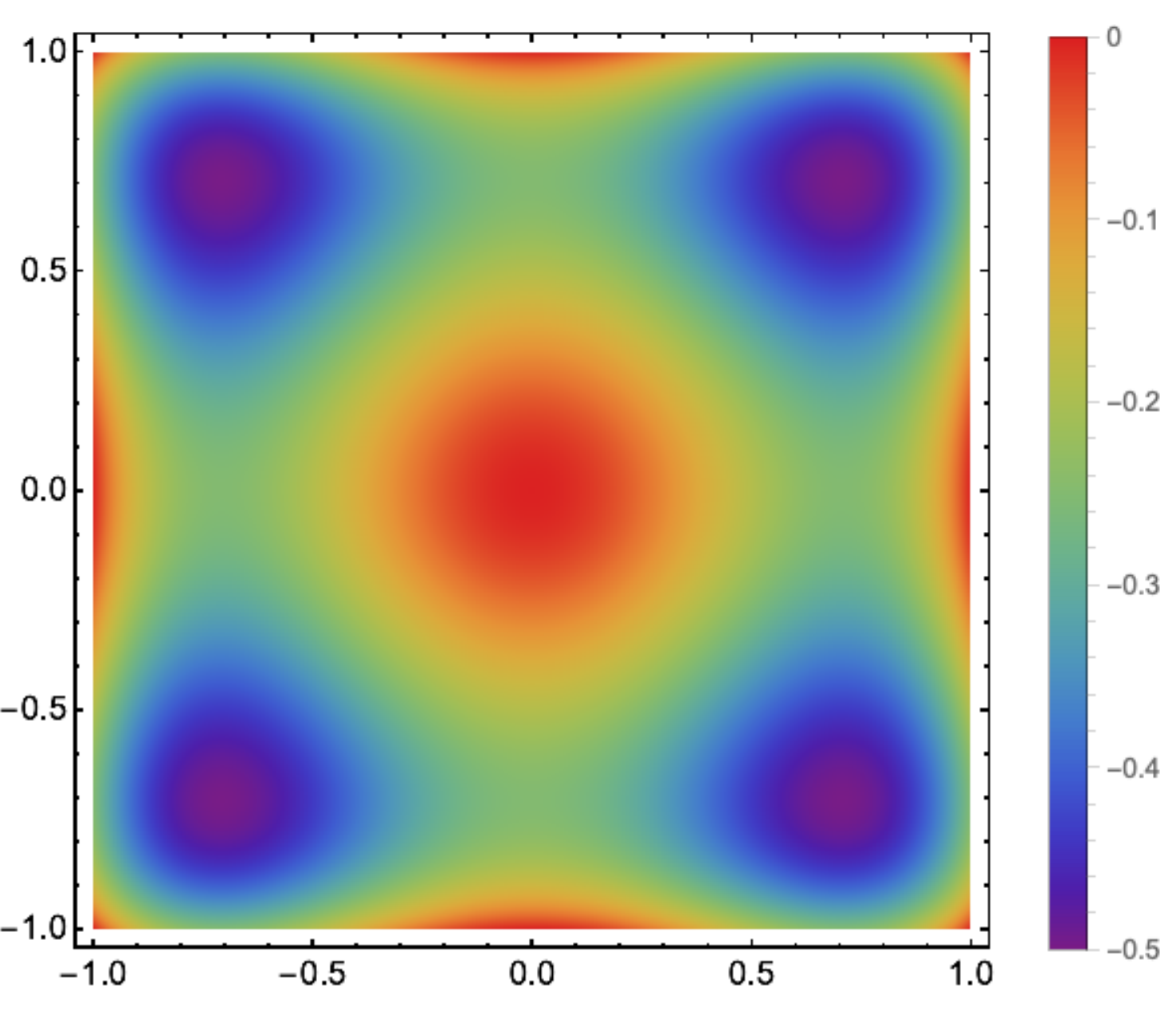}
\caption{Left Panel : The representation of the potential configuration for $\lambda_{12} = 0$. Right panel : the corresponding heat map for such configuration.}
\label{pot3}
\end{figure}

\item \underline{$\lambda_{12} < 0$} : Imposing $\lambda_1+\lambda_2> 0$ as well as $\lambda_{12} < 0$,  $\lambda_2<0$, so we can consider its value to be restricted within $(-1.0,-0.5)$. Here also we obtain 4 distinct minima (distorted and shifted compared to $\lambda_{12}=0$ case) at around $\lbrace \phi_1, \phi_2 \rbrace = (\pm 1, \pm 1)$. The potential structure as well as corresponding heat map in the $\phi_1-\phi_2$ plane are presented in left and right panel of Fig. \ref{pot4} respectively.
\end{enumerate}

A similar analysis for scalar potential part involving only scalar doublets $\chi_{L,R}$ can be carried through by replacing $\mu_1^2 \rightarrow \mu_2^2, \lambda_1 \rightarrow \rho_1, 4\lambda_2 \rightarrow 2(\rho_2 - \rho_1)$ in $\mathcal{V}_{\Phi}$ as discussed in Section \ref{sec:potential}.

\section{Copositivity requirements}
\label{sec:copositivity}
In this section we  study the stability of vacuum structure using Bounded From Below (BFB) and copositivity stabilization criteria. The term copositive refers to "conditionally positive" \cite{Kannike:2016fmd}, and replaces the stronger criterion of positive definiteness of matrices. While its implementation can become complicated, in our case due to only two independent variables to be considered at any time, the criterion becomes much simpler.

\textbf{(i) Boundedness of the potential:} To ensure the stability of the scalar potential one of the criteria to be satisfied  is that the potential should be bounded from below in all field directions in the theory. Specifically, the quartic self-couplings terms in the theory should be positive-definite as field values go to infinity. This criterion is known as Bounded From Below condition.

\textbf{(ii) Copositivity condition:} Given $ax^2+bx+c > 0$ with $x \in \mathbb{R}^+$, copositivity requires that $a > 0, c>0$, and $b+2\sqrt{ac} > 0$.

\subsection{The bidoublet $\Phi$: the $\lambda$ sector}
\label{subsec:bidoublet}

First we investigate the quartic self-interactions of the bidoublet from the potential as described in Eq. \ref{eq:pot}. To satisfy the BFB condition the quartic sector must be positive definite for all values of $\Phi$.
\begin{center}
    $\mathcal{V}_4^\lambda = \lambda_1 \left( \text{Tr} \left[\Phi^{\dagger} \Phi \right] \right)^2
+  \lambda_2 \text{Tr} \left[\tilde{\Phi}^{\dagger}  \Phi \right] \text{Tr} \left[\Phi^{\dagger} \tilde{\Phi} \right] > 0$
\end{center}
To obtain the conditions for BFB, we can parameterize the terms in the potential as
\begin{center}
    $\text{Tr} \left[\Phi^{\dagger} \Phi \right] \equiv r^2$\\
    $\text{Tr} \left[\tilde{\Phi} \Phi^{\dagger}\right] /  \text{Tr} \left[\Phi^{\dagger} \Phi \right]\equiv \xi e^{i \omega}$\\
    $\text{Tr} \left[\tilde{\Phi}^{\dagger} \Phi\right] /  \text{Tr} \left[\Phi^{\dagger} \Phi \right]\equiv \xi e^{-i \omega}$\, ,
\end{center}
where $r>0, \xi \in [0,1]$ and $\omega \in [0,2\pi].$ With this parameterization,
\begin{center}
   $\mathcal{V}_4^\lambda = r^4 (\lambda_1  + \lambda_2 \xi^2) \equiv r^4 f(\lambda, \xi)$ 
\end{center}
For an extremum, the minimum of $\mathcal{V}_4^\lambda$ must exist in/on the closed boundary defined by the disc with radius $\xi$, that is, it  should either exist inside the bounded region or on the boundary of the region. We first minimize the parameterized quartic sector inside the bounded region:
\begin{center}
    $f_\xi \equiv \frac{\partial f}{\partial \xi} =2 \lambda_2 \xi =0 \Rightarrow \xi = 0$
\end{center}
So, from BFB condition for $\xi =0$ we get $\boxed{\lambda_1 >0}$. Now, on the boundary,  setting $\xi =1$ we obtain 
\begin{equation}
    \boxed{\lambda_1 + \lambda_2 > 0$ or $\lambda_1 > - \lambda_2.}
    \label{eq:copositivity_lambda}
\end{equation}
\subsection{The doublet quartic sector involving $\chi_{L}$, $\chi_R$ : the $\rho$ sector}
\label{subsec:doublet}

Similar to the bidoublet case, here we  analyze the quartic sector involving the doublet Higgs only in the potential. In this case
\begin{center}
    $\mathcal{V}_4^\rho = \rho_1 \left[(\chi_L^\dagger \chi_L)^2 + (\chi_R^\dagger \chi_R)\right] + 2\rho_2 (\chi_L^\dagger \chi_L)(\chi_R^\dagger \chi_R)$
\end{center}
Here also, to find the BFB conditions, we  parameterize in similar fashion,
\begin{center}
    $(\chi_L^\dagger \chi_L) + (\chi_R^\dagger \chi_R) \equiv r^2$\\
    $(\chi_L^\dagger \chi_L) \equiv r^2 \sin^2 \gamma$ \\
    $(\chi_R^\dagger \chi_R) \equiv r^2 \cos^2 \gamma$
\end{center}
where $r>0\, , \gamma \in [0,\frac{\pi}{2}]$. As the BFB condition requires that the quartic doublet part of the scalar potential should be positive definite in all $\chi_L, \chi_R$ directions, the corresponding potential, becomes, in the parameterized form 
\begin{align}
    \mathcal{V}_4^\rho &= \rho_1 r^4 \sin^4 \gamma + \rho_1 r^4 \cos^4 \gamma + 2\rho_2 r^4 \sin^2 \gamma \cos^4 \gamma > 0 \nonumber \\
    \Rightarrow & \frac{r^4}{(1+\tan^2 \gamma)^2  } \left[\rho_1 \tan^4 \gamma + 2\rho_2 \tan^2 \gamma +\rho_1\right] > 0
\end{align}
 Imposing the copositivity criteria for $\tan^2 \gamma$, the conditions derived from the doublet sector are
 \begin{equation}
\boxed{\rho_1 > 0, ~~~~~~ \mathcal{G} (\rho) \equiv \rho_2 + \rho_1 > 0\, .}
\label{eq:coupled_cond}
 \end{equation}
 \subsection{The coupled quartic part: the $\alpha$ sector}
 \label{subsec:coupled}
We now  proceed to analyse the complete potential, that is, we include the coupled quartic coupling  involving $\Phi, \chi_{L,R}$ as well as the individual bidoublet and doublet quartic structure), where the total number of free parameters is 4 {\it i.e.}, $\alpha_{1,2,3} \neq 0$ and choose $\mu_3 = 0$\footnote{The mass-dimension parameter $\mu_3$ does not affect the copositivity criterion which put bounds on dimensionless parameters like $\lambda, \alpha, \rho$'s, but it affects the Higgs masses, as will be seen in the next section, Section \ref{sec:masses}.}:
\begin{eqnarray}
&&\mathcal{V}_4  = \lambda_1 \left( \text{Tr} \left[\Phi^{\dagger} \Phi \right] \right)^2
+  \lambda_2 \text{Tr} \left[\tilde{\Phi}^{\dagger}  \Phi \right] \text{Tr} \left[\Phi^{\dagger} \tilde{\Phi} \right]   + \rho_1 \left[ \left(\chi_L^{\dagger} \chi_L \right)^2 +  \left(\chi_R^{\dagger} \chi_R \right)^2 \right]  \nonumber \\
&+& 2 \rho_2 \left(\chi_L^{\dagger} \chi_L\right) \left(\chi_R^{\dagger} \chi_R \right)
+ 2 \alpha_1 \text{Tr} \left[\Phi^{\dagger} \Phi \right] \left(\chi_L^{\dagger} \chi_L +  \chi_R^{\dagger} \chi_R \right)
 +2 \alpha_2 \left[ \chi_L^{\dagger} \Phi \Phi^\dagger \chi_L + \chi_R^{\dagger} \Phi^\dagger \Phi \chi_R \right]  \nonumber \\
 &+ & 2\alpha_3 \left[\chi_L^\dagger \tilde{\Phi} \tilde{\Phi}^\dagger \chi_L + \chi_R^\dagger \tilde{\Phi}^\dagger \tilde{\Phi} \chi_R \right]
\label{eq:V4alpha}
\end{eqnarray}
This $\mathcal{V}_4$ has three field directions. It can be easily shown that $\text{Tr}(\Phi^\dagger \Phi)=\text{Tr}(\tilde{\Phi}^\dagger \tilde{\Phi})$. So here the parameterization can be chosen as follows,
\begin{center}
    $\text{Tr} (\Phi^\dagger \Phi) + \text{Tr} (\tilde{\Phi}^\dagger \tilde{\Phi}) + (\chi_L^\dagger \chi_L) + (\chi_R^\dagger \chi_R) \equiv r^2$ \\
    $\text{Tr} (\tilde{\Phi}^\dagger \tilde{\Phi}) = \text{Tr} (\Phi^\dagger \Phi) \equiv \frac{r^2}{2} \cos^2 \theta$ \\
    $(\chi_L^\dagger \chi_L) \equiv r^2 \sin^2 \theta \sin^2 \gamma$ \\
    $(\chi_R^\dagger \chi_R) \equiv r^2 \sin^2 \theta \cos^2 \gamma$ \\
    $\text{Tr} (\tilde{\Phi}^\dagger \Phi) / \text{Tr} (\Phi^\dagger \Phi) \equiv \xi e^{-i\omega}$ \\
    $\text{Tr} (\Phi^\dagger \tilde{\Phi}) / \text{Tr} (\Phi^\dagger \Phi) \equiv \xi e^{i\omega}$ \\
    $(\chi_L^\dagger \Phi \Phi^\dagger \chi_L) / \text{Tr} (\Phi^\dagger \Phi) (\chi_L^\dagger \chi_L) \equiv \eta_1$ \\
    $(\chi_R^\dagger \Phi^\dagger \Phi \chi_R)/ \text{Tr} (\Phi^\dagger \Phi) (\chi_R^\dagger \chi_R) \equiv \eta_2$\\
    $(\chi_L^\dagger \tilde{\Phi} \tilde{\Phi}^\dagger \chi_L)/\text{Tr} (\tilde{\Phi}^\dagger \tilde{\Phi}) (\chi_L^\dagger \chi_L) = \eta_3 $\\
    $(\chi_R^\dagger \tilde{\Phi}^\dagger \tilde{\Phi} \chi_R)/ \text{Tr} (\tilde{\Phi}^\dagger \tilde{\Phi}) (\chi_R^\dagger \chi_R) = \eta_4\, ,$
\label{eq:polyparameterisation}
\end{center}
where $r > 0$,  $\theta, \gamma \in [0,\frac{\pi}{2}], \xi \in [0,1]$ and $\omega \in [0,2\pi]$ while one would  assume that $\eta_{1,2,3,4}$ can take any value in the interval $[0,1]$, this is not the case. The range allowed for these  parameters  is $\frac{1}{2} (1-\sqrt{1-\xi^2}) < \eta_{1,2,3,4} < \frac{1}{2} (1+\sqrt{1-\xi^2})$ as shown in the derivation  in \cite{Chauhan:2019fji}. Among them $\eta_{1,3}$ and $\eta_{2,4}$ are related to each other as discussed in \cite{Kannike:2021fth} as,
\begin{align}
\eta_1 + \eta_3 = 1, ~~~~~~~~~~~~ \eta_2 + \eta_4 =1.
\label{dependency}
\end{align}

Using the parametric forms defined above we  have
\begin{equation}
    \mathcal{V}_4 = r^4 \left [\cos^4 \theta \, f(\lambda, \xi) + \sin^4 \theta  \, g(\rho,\gamma) + \sin^2 \theta \cos^2 \theta \, h(\alpha, \gamma, \eta_1, \eta_2, \eta_3, \eta_4)\right ]
\end{equation}
where 
\begin{eqnarray}
   && f(\lambda, \xi) \equiv \frac{1}{4}(\lambda_1 + \lambda_2 \xi^2), ~~~~~~~~ g(\rho,\gamma) \equiv \rho_1 \sin^4 \gamma + \rho_1 \cos^4 \gamma + 2\rho_2 \sin^2 \gamma \cos^2 \gamma, \nonumber \\
    &&h(\alpha, \gamma, \eta_1, \eta_2, \eta_3, \eta_4) \equiv \alpha_1 + \alpha_2 (\eta_1 \sin^2 \gamma + \eta_2 \cos^2 \gamma) + \alpha_3 (\eta_3 \sin^2 \gamma + \eta_4 \cos^2 \gamma).
\end{eqnarray}
 The parameterized $\mathcal{V}_4$ yields, using copositivity criteria, 
\begin{eqnarray}
    &&f(\lambda, \xi) > 0, ~~~~~~~~ g(\rho, \gamma) > 0 \nonumber \\
    &&h(\alpha, \gamma, \eta_1, \eta_2, \eta_3, \eta_4) + 2 \sqrt{f(\lambda, \xi)g(\rho, \gamma)} > 0
\end{eqnarray}
Using this third criteria we must have,
\begin{equation}
    \alpha_1 + \alpha_2 (\eta_1 \sin^2 \gamma + \eta_2 \cos^2 \gamma) + \alpha_3 (\eta_3 \sin^2 \gamma + \eta_4 \cos^2 \gamma) + 2\sqrt{f(\lambda, \xi)g(\rho, \gamma)} > 0
\end{equation}
This condition is symmetric under the exchange :  $\eta_1 \leftrightarrow \eta_2, \eta_3 \leftrightarrow \eta_4, \sin \gamma \leftrightarrow \cos \gamma$. 
Its minimum value will be attained inside the proposed gauge orbit space \cite{Kim:1981xu,Kim:1981jj,Frautschi:1981jh,Kim:1983mc,Abud:1981tf,Abud:1983id} \footnote{Gauge orbits are the sets of all possible  gauge configurations of the gauge field in a gauge theory, where various gauge field configurations connected by a single gauge transformation lead to the same physical theory.} when $\eta_1 = \eta_2$, $\eta_3 = \eta_4$ and $\sin \gamma = \cos \gamma = \frac{1}{\sqrt{2}}$. Also $g(\rho, \gamma)$ obeys the symmetry under the exchange : $\sin \gamma \leftrightarrow \cos \gamma$, so it can be minimized using $\sin \gamma = \cos \gamma = \frac{1}{\sqrt{2}}$. Inserting this value we obtain $g_{\text{min}} \equiv \frac{\rho_1 + \rho_2}{2}$ . So the third condition from copositivity criteria can be written as
\begin{equation}
    \alpha_1 + \alpha_2 \eta_1 + \alpha_3 \eta_3 + 2\sqrt{f(\lambda, \xi) \left(\frac{\rho_1 + \rho_2}{2}\right)} > 0
\end{equation}
Now for different relative signs between $\alpha_2$ and $\alpha_3$ we  have $\eta_1 = \eta_2 = \eta_3 = \eta_4 = \eta_{\text{max}} = \frac{1}{2} (1+\sqrt{1-\xi^2})$. Similarly $\eta_1 = \eta_2 = \eta_3 = \eta_4 = \eta_{\text{min}} = \frac{1}{2} (1-\sqrt{1-\xi^2})$. From Eq. \ref{dependency} one can easily infer that $(\eta_1)_{min}=(\eta_3)_{max}$ and vice versa, similar argument applies for $\eta_{2}$ and $\eta_4$. Also $f(\lambda, \xi)$ can be minimized if $\xi = 0$ which translates into $f(\lambda,0) = \lambda_1/4$. So the allowed criteria (after replacing $\eta$'s in terms of $\xi$'s and minimizing the above equation, whole relation can be minimized for $\xi=0$) is translated into
\begin{equation}
    \alpha_1 + \frac{\alpha_2}{2} (1 \pm 1) + \frac{\alpha_3}{2} (1 \mp 1) + \sqrt{\lambda_1 \left(\frac{\rho_1 + \rho_2}{2}\right)} > 0
    \label{1st}
\end{equation}
 We also need to minimize the third condition for the edge surface of $\tan \gamma$. We have already defined the criteria in Eq. \ref{1st} for $\tan \gamma =1$, while $\tan \gamma =0$ is equivalent to $\tan \gamma = \infty$, which results in $g = \rho_1$. So on the boundary we have,  (with $f(\lambda, 0) = \lambda_1/4$), the condition
\begin{equation}
    \alpha_1 + \frac{\alpha_2}{2} (1 \pm 1) + \frac{\alpha_3}{2} (1 \mp 1) + \sqrt{\lambda_1 \rho_1} > 0
\label{2nd}
\end{equation}
Combining the two sets of conditions described in Eqs. \ref{1st} and \ref{2nd} we have the following requirement for the parameters of the complete scalar potential  $\mathcal{V}_4$
\begin{eqnarray}
   && \boxed{\alpha_1 + \alpha_2 + \sqrt{\lambda_1 \left(\frac{\rho_1 + \rho_2}{2}\right)} > 0\, .} \nonumber\\
   && \boxed{\alpha_1 + \alpha_3 + \sqrt{\lambda_1 \left(\frac{\rho_1 + \rho_2}{2}\right)} > 0\, .} \nonumber\\
      \label{eq:vac_stability}
      \\
  &&  \boxed{\alpha_1 + \alpha_2 + \sqrt{\lambda_1 \rho_1} > 0\, .}
   \nonumber \\
   && \boxed{\alpha_1 + \alpha_3 + \sqrt{\lambda_1 \rho_1} > 0\, .}
   \nonumber \\
\end{eqnarray}

\subsection{Symmetry breaking conditions for desirable vacuum}
\label{subsec:ssb}
Finally, in this subsection we investigate the  conditions for correct spontaneous symmetry breaking (SSB) and their effects on the parameters in the scalar potential. The restrictions on the VEVs from the requirement of  symmetry breaking are
\begin{enumerate}
\item \(\langle \Phi \rangle \ne 0, \, ~~\langle \chi_L\rangle \ne  0\)
\item \( \langle \chi_R \rangle \ne 0  \)
\item \(\langle \chi_L\rangle \ne  \langle \chi_R \rangle \)
\end{enumerate}
The first and the second conditions ensure breaking of $SU(2)_L \times U(1)_Y$, and $SU(2)_{R'} \times U(1)_{B-L}$, respectively, whereas the last condition is required to ensure the correct hierarchy of breaking scales.
Inserting the parameterizations of the VEVs Eq. \ref{eq:VEVs} in $\mathcal{V}_4$, the  potential 
becomes
\begin{align}
\tilde{\mathcal{V}}_{SSB} = & \frac{\lambda_1}{4} (v_1^2 + v_2^2)^2 + \lambda_2 v_1^2 v_2^2 + \frac{\rho_1}{4} (v_L^4 +v_R^4) + \frac{\rho_2}{2} v_L^2 v_R^2 + \frac{\alpha_1}{2} (v_1^2 + v_2^2) (v_L^2 + v_R^2) + \frac{\alpha_2 v_2^2}{2} (v_L^2 + v_R^2) \nonumber \\
&  +\frac{\alpha_3 v_1^2}{2} (v_L^2 + v_R^2)
\label{eq:SSB}
\end{align}
For convenience we can adopt the same parameterisation for the gauge invariant monomials as in Section \ref{eq:polyparameterisation} below Eq. \ref{eq:V4alpha}, now in terms of the VEVs which gives,
\begin{align}
&v_1^2 + v_2^2 + \frac{1}{2}(v_L^2 + v_R^2) = r^2 \nonumber \\
&v_1^2 + v_2^2 = r^2 cos^2 \theta \nonumber \\
&v_L^2 = 2 r^2 sin^2 \theta sin^2 \gamma \nonumber \\
&v_R^2 = 2 r^2 sin^2 \theta cos^2 \gamma \nonumber \\
&v_1 v_2 = \frac{r^2 \xi cos^2 \theta}{2} \nonumber \\
&v_1^2 = \eta_3 r^2 cos^2 \theta = \eta_4 r^2 cos^2 \theta \nonumber \\
  &v_2^2 = \eta_1 r^2 cos^2 \theta = \eta_2 r^2 cos^2 \theta
\end{align}
Using these we get
\begin{eqnarray}
\mathcal{V}_{\rm SSB}&=& \lambda_1 \frac{r^4}{4} \cos^4 \theta + \lambda_2 \xi^2 \frac{r^4}{4} \cos^4 \theta +\rho_1 r^4 \sin^4 \theta \sin^4 \gamma +\rho_1 r^4 \sin^4 \theta \cos^4 \gamma + 2 \rho_2  r^4 \sin^4 \theta \sin^2 \gamma \cos^2 \gamma\nonumber \\
&+ &   \alpha_1 r^4 \sin^2 \theta \cos^2 \theta + \alpha_2 \eta_1 r^4 \sin^2 \theta \cos^2 \theta \sin^2 \gamma + \alpha_2 \eta_2 r^4 \sin^2 \theta \cos^2 \theta \cos^2 \gamma
  \nonumber \\
&+&  \alpha_3 \eta_3 r^4 \sin^2 \theta \cos^2 \theta \sin^2 \gamma + \alpha_3 \eta_4 r^4 \sin^2 \theta \cos^2 \theta \cos^2 \gamma  \, 
\label{eq:SSBparameterised}
\end{eqnarray}
Factorised in powers of $\cos\theta$ and $\sin\theta$, this gives
\begin{equation}
\mathcal{V}_{\rm SSB}= r^4\left [ f_{\rm SSB}(\lambda, \xi) \cos^4 \theta+g_{\rm SSB} \sin^4 \theta+h_{\rm SSB} \sin^2 \theta \cos^2\theta\right] \, ,
\end{equation}
where we defined the values of the functions after SSB
\begin{eqnarray}
f_{\rm SSB}(\lambda, \xi) &=& \frac{\lambda_1}{4}+\frac{\lambda_2}{4} \xi^2\nonumber \\
g_{\rm SSB} (\rho, \gamma)&=& \rho_1 \sin^4 \gamma +\rho_1 \cos^4 \gamma +2 \rho_2 \cos^2 \gamma \sin^2 \gamma \nonumber \\
h_{\rm SSB}(\alpha, \gamma, \eta)&=& \alpha_1 + \alpha_2 \left ( \eta_1 \sin^2 \gamma +\eta_2 \cos^2 \gamma \right) + \alpha_3  \left ( \eta_3 \sin^2 \gamma +\eta_4 \cos^2 \gamma \right) 
\end{eqnarray}
Notice that we have considered the general vacuum structure without restricting $v_1$ to vanish. In this new set of parametrisation, setting $v_1$ to zero is consistent with setting $\xi=0=\eta_3=\eta_4$. The copositivity conditions derived below can be considered with the corresponding expressions duly simplified. 

The necessary and sufficient conditions for correct symmetry breaking require the minimum of $\mathcal {V}_{\rm SSB}$ to be deeper than that obtained from the general potential \footnote{A detailed discussion on gauge orbit spaces and its connection to find out electromagnetic charge-preserving vacuum of theory can be found in \cite{Chauhan:2019fji, Kim:1981xu, Kim:1981jj, Frautschi:1981jh, Kim:1983mc, Abud:1981tf, Abud:1983id}).}
$$-\frac{g\mu_1^4-h \mu_1^2 \mu_2^2+f \mu_2^2}{4 fg-h^2} >-\frac{g_{\rm SSB}\mu_1^4-h_{\rm SSB} \mu_1^2 \mu_2^2+f_{\rm SSB} \mu_2^2}{4 f_{\rm SSB}g_{\rm SSB}-h^2_{\rm SSB}}$$
For our potential $g=g_{\rm SSB}, f=f_{\rm SSB}, h=h_{\rm SSB}$ and the minimum of the potential $\mathcal{V}_{\rm SSB}$ is the same as for $\mathcal{V}_4$.
Imposing $\mathcal {V}_{4} \ge \mathcal {V}_{\rm SSB}$ the VEV structure yields, for the global minimum of the theory
$$ f \ge f_{\rm SSB}, \qquad g \ge g_{\rm SSB}, \qquad h - h_{\rm SSB} +2 \sqrt{(f - f_{\rm SSB})( g - g_{\rm SSB})} \ge 0.$$
In addition $\mathcal{V}_{\rm SSB}$ exhibits a stable vacuum for 
$$f_{\rm SSB} >0, \qquad g_{\rm SSB}>0, \qquad h_{\rm SSB}+2 \sqrt{f_{\rm SSB}g_{\rm SSB}}>0.$$
The VEV condition $\langle \Phi \rangle \ne 0$ is satisfied as long as the $\lambda$'s are bounded from below, and all conditions found for charge-preserving vacua trivially satisfy the requirements for the symmetry breaking. The function
$$g_{\rm SSB}=\frac{1}{(1+\tan^2 \gamma)^2} [\rho_1 \tan^4 \gamma +2 \rho_2 \tan^2 \gamma+\rho_1]$$
has a minimum for $\tan^2 \gamma=0$ and 1. 
Imposing further the second SSB condition,  $\langle \chi_L\rangle \le \langle \chi_R \rangle$, $\tan^2 \gamma=0$ is preferred and 
\begin{equation}
\frac{2 \rho_1+2 \rho_2}{4} \ge \rho_1\, ,
\end{equation}
yielding
 \begin{equation}
\boxed{\rho_2 -3 \rho_1 \ge 0\, .}
 \end{equation}
As the minimum of $g_{\rm SSB}$ occurs for $\rho_1>0$, the above condition also implies that $\rho_2>0$. Thus in addition to the constraint  Eq. \ref{eq:coupled_cond}, spontaneous symmetry breaking conditions also imposes $\rho_2>0$ on the parameters of the scalar potential.

\section{Restrictions on the parameter space due to Higgs masses}
\label{sec:masses}
In addition to the vacuum stability conditions analyzed in the previous section, the scalar potential is bound by the requirement of non-tachyonic Higgs boson masses. To insure these conditions are satisfied, we recall the expressions for the masses as given in \cite{Frank:2019nid}.

In the model, based on the generalized lepton number $L=S+T_{3R}$, one introduces a generalized $R$ parity, similar to the one introduced in supersymmetry \cite{Khalil:2010yt}, $(-1)^{3B+2j+L}$, under which all SM quarks and leptons are even, while in the scalar sector, $\chi_R^\pm, \, \phi_1^\pm, \, \Re(\phi_1^0)$ and $\Im(\phi_1^0)$ are odd, while the rest of the Higgs bosons are even. 

This is reflected in the mixing matrices, which are consistent with $R$-parity even and $R$-parity odd scalars not mixing. Indeed in the charged scalar sector, the squared mass matrix is block diagonal. The $\phi^\pm_2$ and $\chi_L^\pm$ ($R$-parity even) fields  mix  independently of the $\phi_1^\pm$ and $\chi_R^\pm$ ($R$-parity odd) fields. The  $2\times 2$ block  mass matrices
$({\cal M}^\pm_L)^2$ and $({\cal M}^\pm_R)^2$ are written, respectively, in
the $(\phi_2^\pm, \chi_L^\pm)$ and $(\phi_1^\pm, \chi_R^\pm)$ bases, as
\renewcommand{\arraystretch}{1.4}
\begin{eqnarray}
  ({\cal M}^\pm_{L,R})^2 = \begin{pmatrix}
     -(\alpha_2-\alpha_3)v_{\scriptscriptstyle L,R}^2 -
         \frac{\mu_3 v_L v_R}{\sqrt{2}v}~~~~~ & 
     (\alpha_2-\alpha_3) v_2 v_{\scriptscriptstyle L,R} +
         \frac{\mu_3 v_{\scriptscriptstyle R,L}}{\sqrt{2}}\\
     (\alpha_2-\alpha_3) v + \frac{\mu_3 v_{\scriptscriptstyle R,L}}{\sqrt{2}}&
     -(\alpha_2-\alpha_3)v^2 - \frac{\mu_3 v v_{\scriptscriptstyle R,L}}
       {\sqrt{2}v_{\scriptscriptstyle L,R}}\end{pmatrix} \ ,
\end{eqnarray}
The masses of the charged Higgs bosons  are obtained simply by diagonalising a $2 \times 2$ matrices. The masses for the charged Higgs bosons are 
\begin{eqnarray}
    m^2_{h_1^{\pm}} &=& - \left[ v_2 v_L \left( \alpha_2 - \alpha_3 \right) + \frac{\mu_3 v_R}{\sqrt{2}}\right]\frac{v^2}{v_2 v_L}\\
    \label{charged1}
    m^2_{h_2^{\pm}} &= &- \left[ v_2 v_R \left( \alpha_2 - \alpha_3 \right) + \frac{\mu_3 v_L}{\sqrt{2}}\right]\frac{v^{\prime 2}}{v_2 v_R}
    \label{charged2}
\end{eqnarray}
with $v^2 = v_2^2 + v_L^2$ and $v^{\prime 2} = v_2^2 + v_R^2$, and where $h_1$ is $R$-parity odd and $h_2$ is $R$-parity even.
The other two eigenstates of these matrices correspond to the Goldstone bosons $G_2^\pm$ ($R$-parity even) responsible for giving mass to the $W_L^\pm$ boson, while the Goldstone $G_1^\pm$ ($R$-parity odd) gives mass to $W_R^\pm$, which is also odd under $R$-parity.

In the pseudoscalar and  scalar sector, components of the $\phi_1^0$ field ($\Re(\phi_1^0)$ and $\Im(\phi_1^0)$)  do not mix with other states, as they are both $R$-parity odd, and they yield  the  physical $h_1^0$ and $A_1$ eigenstates. They are mass-degenerate, with masses  $m_{h_1^0}$ and $m_{A_1}$.

The mass for the CP-odd {and $R$-parity odd) Higgs boson $A_1$ is
\begin{eqnarray}
    m_{A_1}^2 &=& 2 v_2^2 \lambda_2 - \left( \alpha_2 - \alpha_3 \right) \left(v_L^2 + v_R^2 \right) - \frac{\mu_3 v_L v_R}{\sqrt{2}v_2}
    \label{odd1}
    \end{eqnarray}
and the mass for the CP-even, $R$-parity odd Higgs boson $\Re(\phi_1^0)$ is
    \begin{eqnarray}
    m_{h_1^0}^2 &=& m_{A_1}^2
     \label{even1}
      \end{eqnarray}
  
The squared mass matrices $({\cal M}^0_\Re)^2$ and $({\cal M}^0_\Im)^2$ of the
three remaining scalar and pseudoscalar fields (all of which are $R$-parity even) are respectively given, in the
$(\Re\{\phi^0_2\}, \Re\{\chi_L^0\}, \Re\{\chi_R^0\})$ and
$(\Im\{\phi^0_2\}, \Im\{\chi_L^0\}, \Im\{\chi_R^0\})$ bases, by the matrices
\renewcommand{\arraystretch}{1.3}
\setlength\arraycolsep{5pt}
\begin{eqnarray}
  ({\cal M}^0_\Re)^2 &=&\ \begin{pmatrix}
    2v^2 \lambda_1 \!-\! \frac{\mu_3 v_L v_R}{\sqrt{2} v} & 2 \alpha_{12} v v_L
      \!+\! \frac{\mu_3 v_R}{\sqrt{2}} & 2 \alpha_{12} v v_R \!+\!
      \frac{\mu_3 v_L}{\sqrt{2}}\\
    2\alpha_{12} v v_L \!+\! \frac{\mu_3 v_R}{\sqrt{2}} & 2 \rho_1 v_L^2
      \!-\!\frac{\mu_3 v v_R}{\sqrt{2} v_L} & 2 \rho_1 v_L v_R\!-\!
      \frac{\mu_3 v}{\sqrt{2}}\\
    2\alpha_{12} v v_R \!+\! \frac{\mu_3 v_L}{\sqrt{2}} & 2\rho_1 v_L v_R\!-\!
       \frac{\mu_3 v}{\sqrt{2}}  & 2 \rho_1 v_R^2 \!-\!
       \frac{\mu_3 v v_L}{\sqrt{2} v_R}
  \end{pmatrix} \ , \\
  ({\cal M}^0_\Im)^2 &=&\ \frac{\mu_3}{\sqrt{2}} \begin{pmatrix}
    -\frac{v_L v_R}{v} & v_R & - v_L \\
    v_R & -\frac{v v_R}{v_L} & v\\
    -v_L & k & -\frac{v v_L}{v_R} \end{pmatrix} \ ,
\end{eqnarray}
Here the pseudoscalar boson $A_2$ has mass
  \begin{eqnarray}
    m_{A_2}^2 &=&  - \frac{\mu_3 v_L v_R}{\sqrt{2}v_2} \left[ 1+ v_2^2 \left(\frac{1}{v_L^2} + \frac{1}{v_R^2} \right)\right] \, .
    \label{odd2}
\end{eqnarray}
The requirement that CP-odd physical scalar $A_2$ mass be non-tachyonic constrains $\mu_3$ to be negative, as all the other parameters in Eq. \ref{odd2} are positive. The other two CP-odd states  in  
$({\cal M}^0_\Im)^2$ are Goldstone bosons $G_1^0$ and $G_2^0$  for $Z,\,  Z^\prime$ gauge bosons. Both the Goldstone bosons and neutral gauge bosons are $R$-parity even, thus conserving $R$-parity.

The masses for CP-even, $R$-parity even Higgs are $m_{h^0_{0,2,3}}$, where $h^0_0$ corresponds to the SM Higgs.
It is conventional to fix the $\lambda_1$  parameter of the scalar
potential in terms the mass of the lightest Higgs state $h_0^0$ (that can then be set
to match the SM Higgs boson mass). With this, $\lambda_1$ becomes formally a
parameter dependent on $m_{h_0^0}$,
\begin{eqnarray}
\label{eq:lam1}
  \lambda_1 =  \frac{1}{2 v^3}\ \frac{
    \sqrt{2} v v_L v_R m_{h_0^0}^6 + \mathfrak{a}^{(4)} m_{h_0^0}^4
      - 2\mathfrak{a}^{(2)}m_{h_0^0}^2 - 4\alpha_{12}^2\mu_3 v^4(v_L^2-v_R^2)^2
  }{
    \sqrt{2} v_L v_R m_{h_0^0}^4
      + (\mu_3 v - 2\sqrt{2}\rho_1 v_L v_R) (v_L^2+v_R^2)m_{h_0^0}^2 
      - 2\mu_3 v \rho_1 (v_L^2-v_R^2)^2
  }\ .
\end{eqnarray}
However, we found that, since $\lambda_1$ depends on several other parameters ($v_l, \, v_R,  \,v, \,  \rho_1,  \, \mu_3,  \,\alpha_{12}$ and $\mathfrak{a}$), it is simpler to vary it and study the effects on the vacuum state.
The rest of the CP-even ($R$-parity-even) neutral Higgs boson masses are
    \begin{eqnarray}
    m_{h^0_{2,3}}^2 &=& \frac{1}{2} \left [\mathfrak{a} - m_{h}^2 \mp \sqrt{\left(\mathfrak{a} - m_{h}^2 \right)^2 + 4 \left(\mathfrak{b} + m_{h}^2 \left(\mathfrak{a} - m_{h}^2 \right) \right) }\right]\,,
    \label{even2}
\end{eqnarray}
where 
\begin{eqnarray}
    \mathfrak{a} &= &2 v_2^2 \lambda_1 + 2 \left(v_L^2 + v_R^2 \right)\rho_1 - \frac{\left( v_L^2 v_R^2 + v_2^2 \left[v_L^2 + v_R^2 \right)\right] \mu_3}{\sqrt{2}v_2 v_L v_R}, \nonumber \\
    \mathfrak{b} &=& \frac{\left( v_L^2 + v_R^2 \right)}{v_2 v_L v_R} \left\{ 4 v_2^3 v_L v_R \left[ \left( \alpha_1 + \alpha_2 \right)^2 -  \lambda_1 \rho_1  \right] + \sqrt{2}v_2^4 \lambda_1 \mu_3 + \sqrt{2}v_L^2 v_R^2 \rho_1 \mu_3 \right\} \nonumber \\
    &+& \frac{\sqrt{2}v_2 \mu_3}{v_L v_R} \left[ 4 v_L^2 v_R^2 \left( \alpha_1 + \alpha_2 \right) + \left(v_L^2 - v_R^2 \right)^2 \rho_1 \right]. \nonumber 
\end{eqnarray}

The existence of an $R$-parity odd sector, which mixes among itself only, and thus acts as like separate sector of the model, raises the question of whether the lightest $R$-parity odd Higgs particle could serve as a dark matter candidate. While a full investigation of this does not exist in the literature, the scotino, which is also odd under $R$-parity, has been shown to be a promising dark matter candidate for the model \cite{Frank:2019nid}. An analysis of the possibility that one of the Higgs states could be a consistent dark matter involves a separate analysis of the parameter space required to satisfy annihilation and co-annihilation cross sections yielding correct relic density and satisfying direct detection constraints. This is beyond the scope of the present paper, and we leave it for investigation in future work.

Imposing $\mu_3 <0$ as required by avoiding tachyonic masses, Eq. (\ref{odd2}),  we distinguish two different cases :
\begin{itemize}
\item(a) For $\left( \alpha_2 - \alpha_3 \right) \ge 0$,  $m^2_{h_1^{\pm}} >0$ but $m^2_{h_2^{\pm}} <0$.
\item(b) For $\left( \alpha_2 - \alpha_3 \right) \le 0$, $m^2_{{h_1^{\pm}} ,{h_2^{\pm}}} >0$.
\end{itemize}
These further constrain  $(\alpha_2 - \alpha_3)$ as $(\alpha_2 - \alpha_3) \le 0$. But from the copositivity criterion puts constraint on $(\alpha_2 - \alpha_3)$ as $(\alpha_2 - \alpha_3) \ge 0$. So, it appears that only the allowed parameter space is the one where 
$\alpha_2 = \alpha_3. $

We  proceed to investigate the parameter space more carefully by imposing not only the non-tachyonic conditions on the mass values, but also the requirement that they satisfy the mass bounds for additional Higgs bosons as in \cite{Zyla:2020zbs}. The charged scalar $h_2^\pm$ and the pseudo-scalar $A_1$ are right-handed and decay into an ordinary and an exotic fermion. Assuming they are light, these particles are long lived but are not produced in $q \bar{ q}^{(')}$, and thus their masses are not bounded by collider data. In contradistinction, $h_2^\pm$, and $A_2$ decay into ordinary fermions only and can be produced in Drell-Yan  processes, and thus their masses are restricted by searches of new Higgs bosons. In Fig. \ref{fig:Higgs} we plot the restrictions on the parameters $\mu_3$ and $\alpha_2-\alpha_3$ coming from these constraints.
\begin{figure}[htb!]
\centering
\includegraphics[scale=0.55]{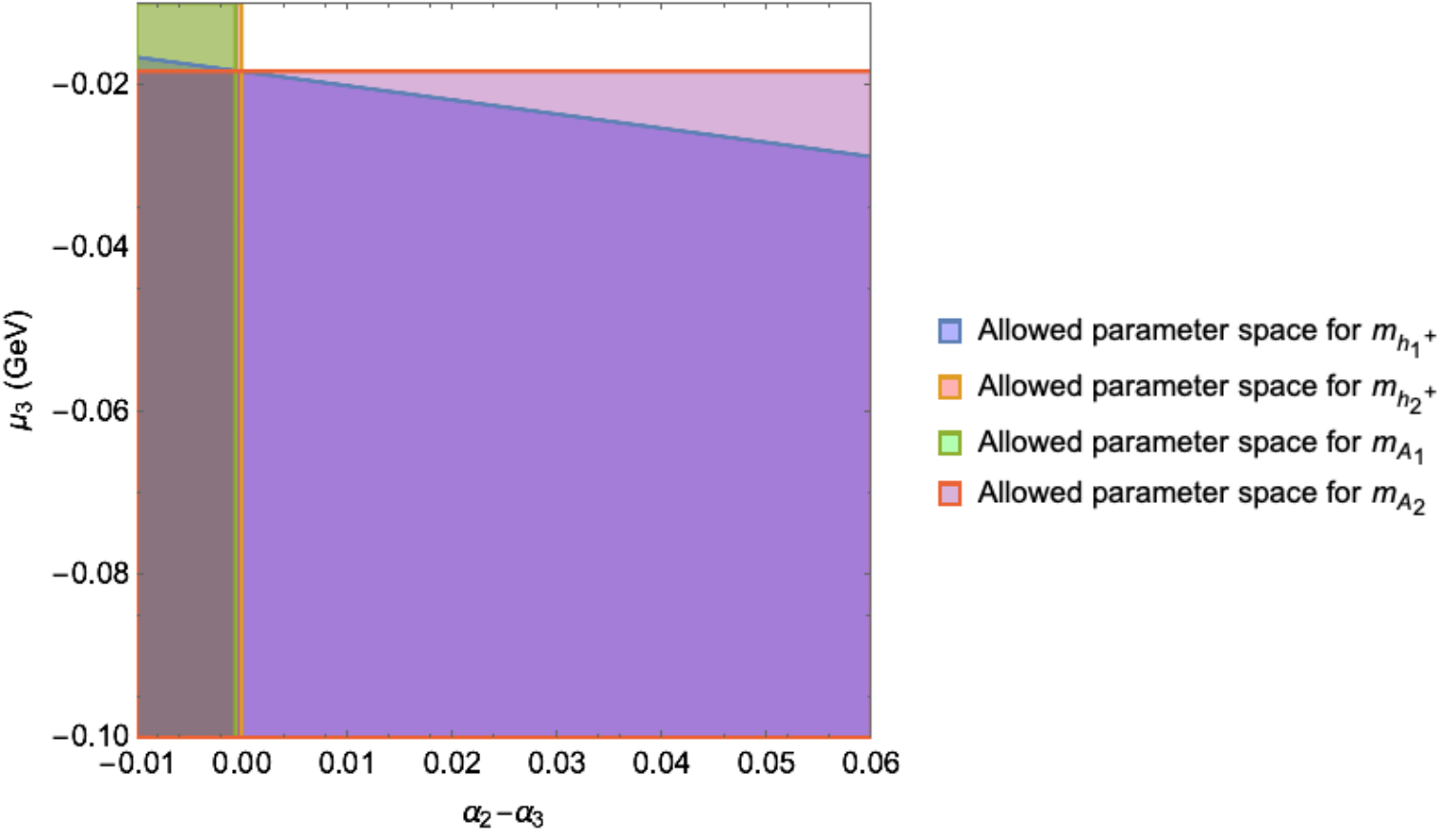}\qquad
\includegraphics[scale=0.45]{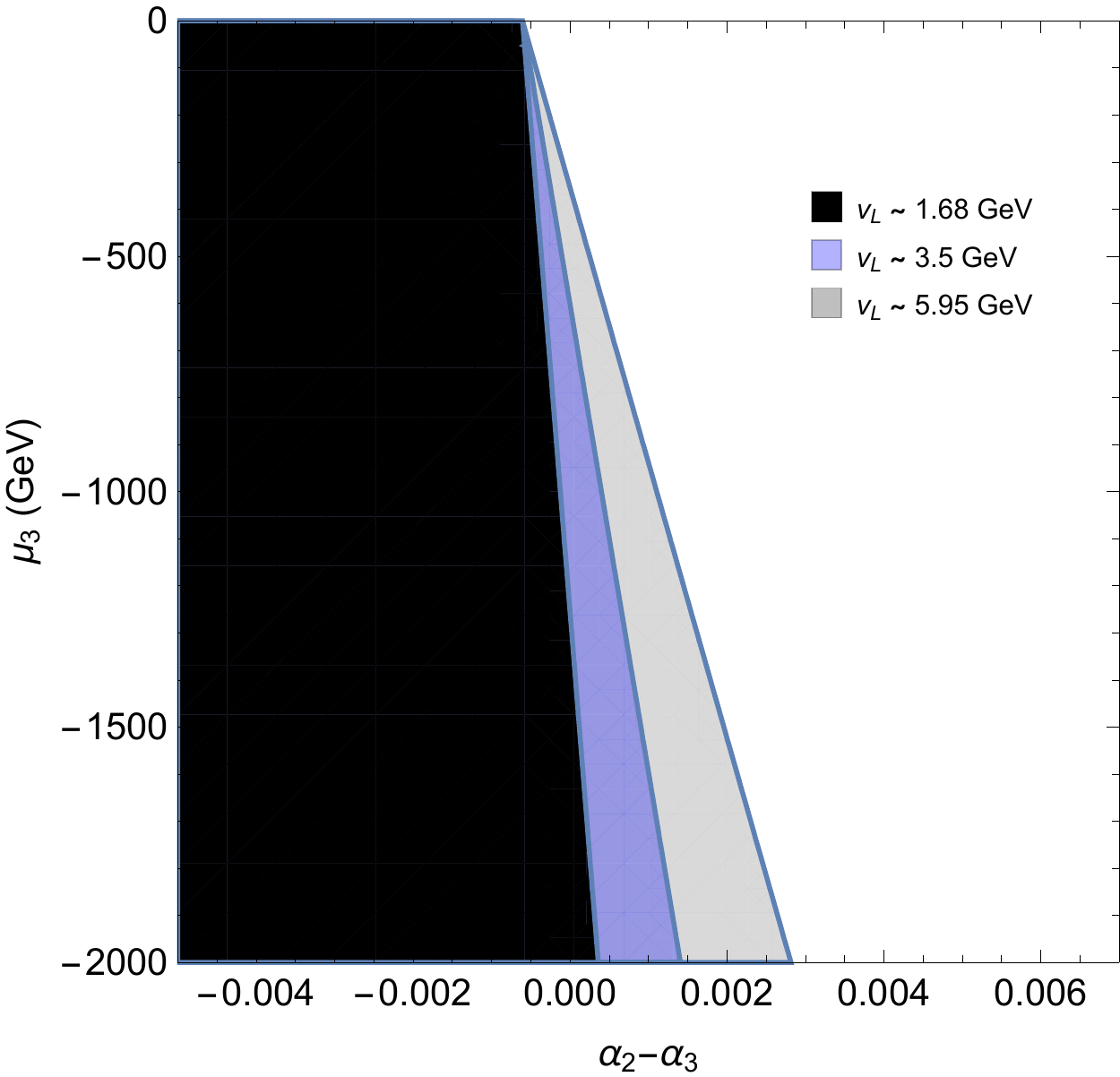}
\caption{Parameter space spanned by $\mu_3$ and $\alpha_2-\alpha_3$. Left Panel : Allowed regions (shaded) are obtained considering restrictions on the physical charged and neutral scalar states, requiring $m_{h_1^\pm} > 80$ GeV, $m_{h_2^\pm}, m_{A_1} > 0$ and $m_{A_2} > 80$ GeV \cite{Zyla:2020zbs}. Right Panel : Surviving allowed  parameter space from Higgs mass criteria for different values of $v_L$, as given in the legend.}
\label{fig:Higgs}
\end{figure}

In the left panel of Fig. \ref{fig:Higgs}, we plotted several contributions arising due to Higgs mass relations (quoted in Eqs. \ref{charged1}- \ref{odd2}) to constrain the parameter space between $\mu_3$ and $(\alpha_2-\alpha_3)$. In this figure, the blue region describes the parameter space allowed by charged scalar mass $m_{h_1^\pm} > 80$ GeV, while the second CP-odd Higgs $A_2^0$ mass  ($m_{A_2} > 80$ GeV) constraints the parameter space within the rectangular region as indicated in red. Also from the allowed Yukawa couplings in ALRM framework, charged Higgs $H_2^\pm$ and CP-odd scalar $A_1^0$ can decay to exotic fermions thus we only require their masses  be positive definite. Using the conditions $m_{h_2^\pm}, m_{A_1} > 0$, we constrained the parameter space in the left panel of Fig. \ref{fig:Higgs} represented by orange and green rectangular regions. In the plots we keep $v_R=10$ TeV constant, as required by the $Z^\prime$ mass lower bounds, but vary $v_L$. All these constraints coming from Higgs mass consideration impose stronger limits on the parameter space $\mu_3$ \textit{vs} $\alpha_2-\alpha_3$. In the right panel of the figure, the overlapped regions obtained from all the  shaded region shows which clearly  that $\mu_3 < - 353.54~ (-601.23, -1253.17)$ GeV (or $|\mu_3| > 353.54~ (601.23, 1253.17)$ GeV for $v_L=5.96~ (3.5, 1.68) $ GeV, as well as $(\alpha_2-\alpha_3) = 0$. From our plot presentation, we show $\mu_3 \in [-0.001,-0.1]$ GeV in left panel of the figure but $\mu_3$ can take further values along the negative y-axis, where we show for different $v_L$ values in right panel of the figure. 

\section{Higgs mass dependence on the parameters}
\label{sec:higgs}
In this section, we analyze the mass dependence of the scalar, pseudo-scalar and charged Higgs bosons as a function of the relevant parameters in the scalar potential.  Our aim is to show the effects of requiring the parameter space to be consistent with the vacuum structure and look for viable Higgs masses. The advantage of this model is that there are no FCNC in the Higgs sector,  thus the scalar masses can be light, and the corresponding Higgs bosons could observed at the LHC.

First, we analyze the mass dependence of the pseudo-scalar and charged Higgs states. The expressions for these masses are quite simple, Eqs.(\ref{charged1}- \ref{odd2}), and for fixed $v_2, v_L$ and $v_R$ they depend on $\mu_3$ only. The dependence is shown in Fig.  \ref{fig:mu3_dep}.
\begin{figure}
\centering
 \includegraphics[scale=0.7]{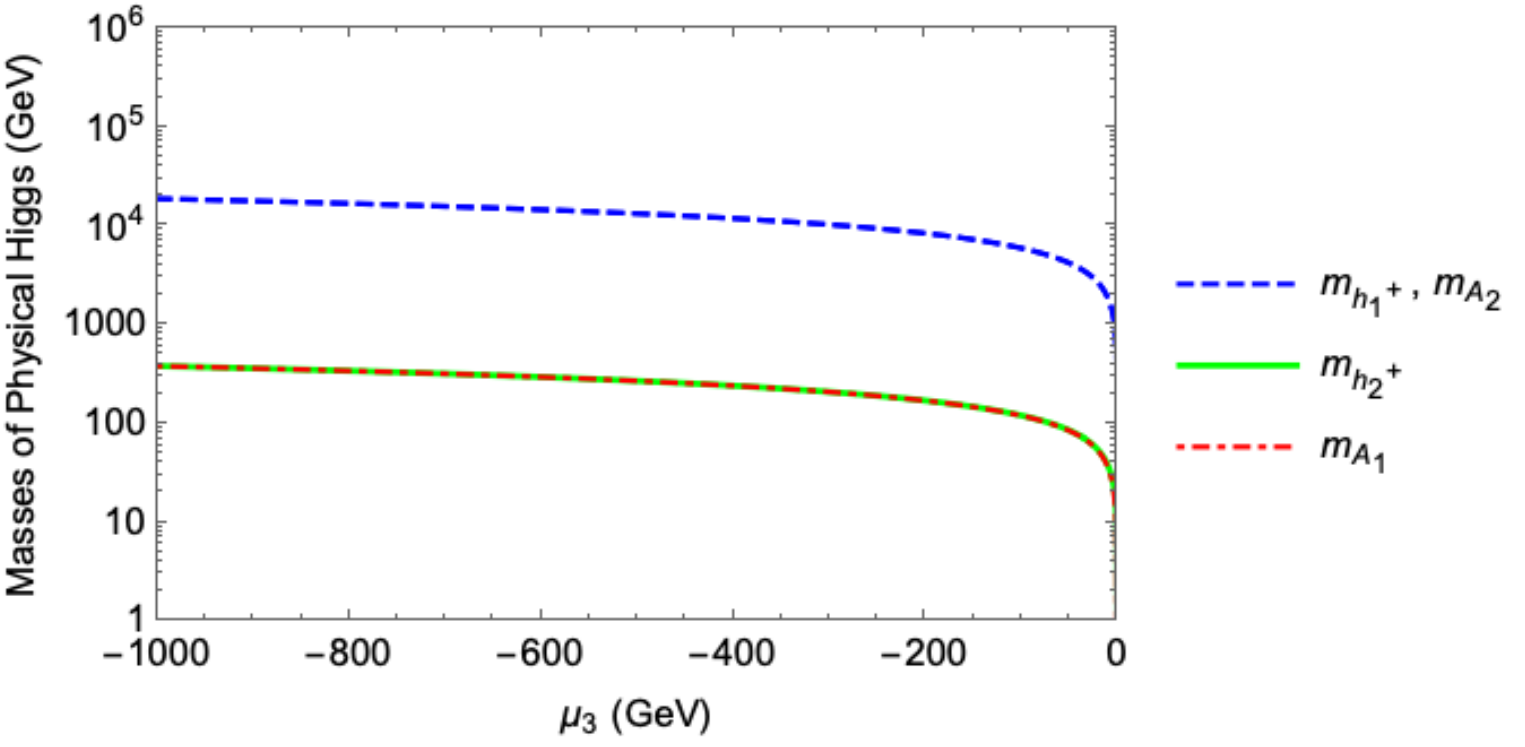} 
\caption{Charged  and pseudo-scalar Higgs boson masses as  functions of $\mu_3$.}
\label{fig:mu3_dep}
\end{figure}

Next we investigate the parameter dependence of the CP-even Higgs bosons. The expressions are more complicated, Eqs.(\ref{even1}- \ref{even2}). These masses depend on $\mu_3$ and in addition, on the sum $\alpha_1+\alpha_2$, $\rho_1$ and $\lambda_1$. We show, in Fig. \ref{fig:evenmu3_dep} the variation of $m_{h_2}$ and $m_{h_3}$ with $\mu_3$ for some choices for $\alpha_1, \alpha_2, \rho_1$ and $\lambda_1$. 
For either light or heavy CP-even Higgs masses, $m_{h_2}$ is independent of $\mu_3$, except in the region $|\mu_3|<1$ GeV, when the mass drops abruptly (due to the condition of requiring $\mu_3<0$ for non-tachyonic masses. Whereas $m_{h_3}$ decreases when we decrease $|\mu_3|$, but is always positive, Fig. \ref{fig:evenmu3_dep}. However, there is strong dependence of the $\lambda_1\, , \rho_1$ parameters as seen from Eq. \ref{even2}., showing that, with the definitions of $a$ and $b$ parameters, $m_{h_2}, m_{h_3}  \approx \sqrt{\lambda_1}\, , \sqrt{\rho_1}$.

In Fig. \ref{fig:alpha_dep}  we plot the variation of the masses of the CP-even Higgs $h_2$ and $h_3$ with $\alpha_1+\alpha_2$, for some choices for the $\mu_3$ parameters, and with $\lambda_1$ and $\rho_1$, chosen to be equal. From the derived results we can see $(\alpha_1+\alpha_2)^2 > \lambda_1 \rho_1$, so we have chosen as benchmarks $\lambda_1 = \rho_1 = 0.99 (\alpha_1 + \alpha_2)$, $\lambda_1 = \rho_1 = 0.7 (\alpha_1 + \alpha_2)$, $\lambda_1 = \rho_1 = 0.5 (\alpha_1 + \alpha_2)$, and $\lambda_1 = \rho_1 = 0.3 (\alpha_1 + \alpha_2)$.

 We show masses of $h_2$  in the left panels, and $h_3$ in the right ones. Choosing different $\lambda_1 = \rho_1$ to be smaller multiples of $\alpha_1 + \alpha_2$ lowers the masses of both $h_2$ and $h_3$, but the dependence is different: while $m_{h_2}$ increases linearly with $\alpha_1 + \alpha_2$ up to 0.05, where it reaches a plateau and then decreases, $m_{h_3}$ is constant when increasing $\alpha_1 + \alpha_2$ up to 0.05, and then it increases linearly. Throughout the parameter range, $m_{h_2}$ remains relatively light, below 800 GeV, and for $\alpha_1 + \alpha_2 \le 0.05$ , $m_{h_3}$ is below 1 TeV. 
 
 Looking at the dependence on $\lambda_1=\rho_1$ for two values of  $\mu_3$, lighter masses for $h_2$ are obtained for larger $\lambda_1=\rho_1$, while  $m_{h_3}$ increases  parabolically with $\mu_3$ for smaller $|\mu_3|$ values and linearly for larger $|\mu_3|$ values up to  $\lambda_1=\rho_1=3$, where it becomes independent of $\mu_3$. 
 
\begin{figure}
\centering
 \includegraphics[scale=0.65]{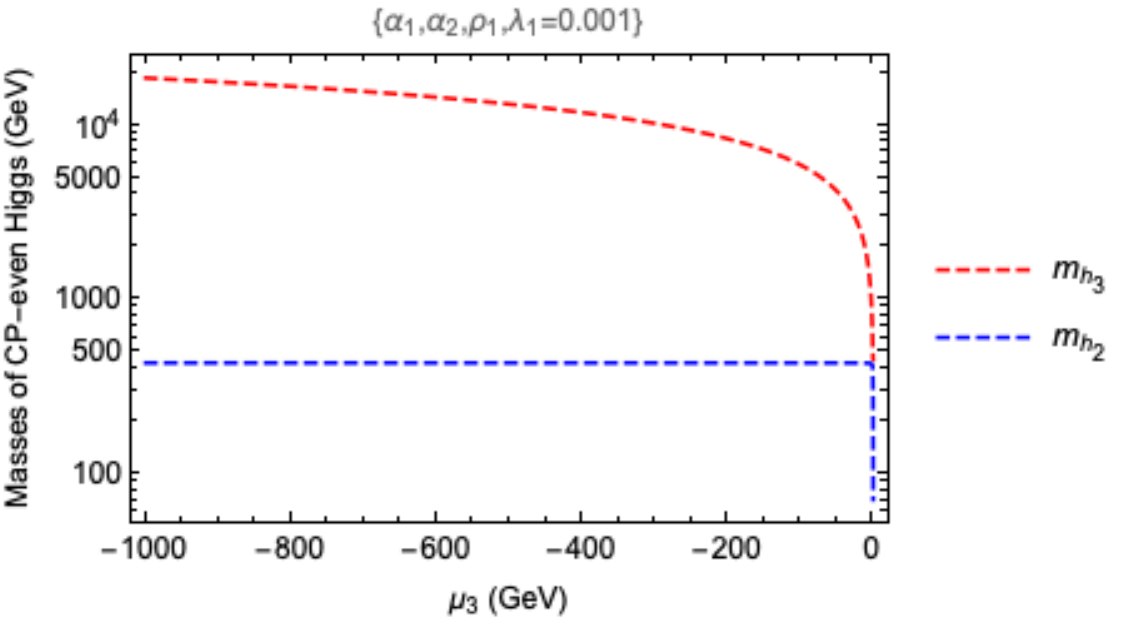} 
\includegraphics[scale=0.65]{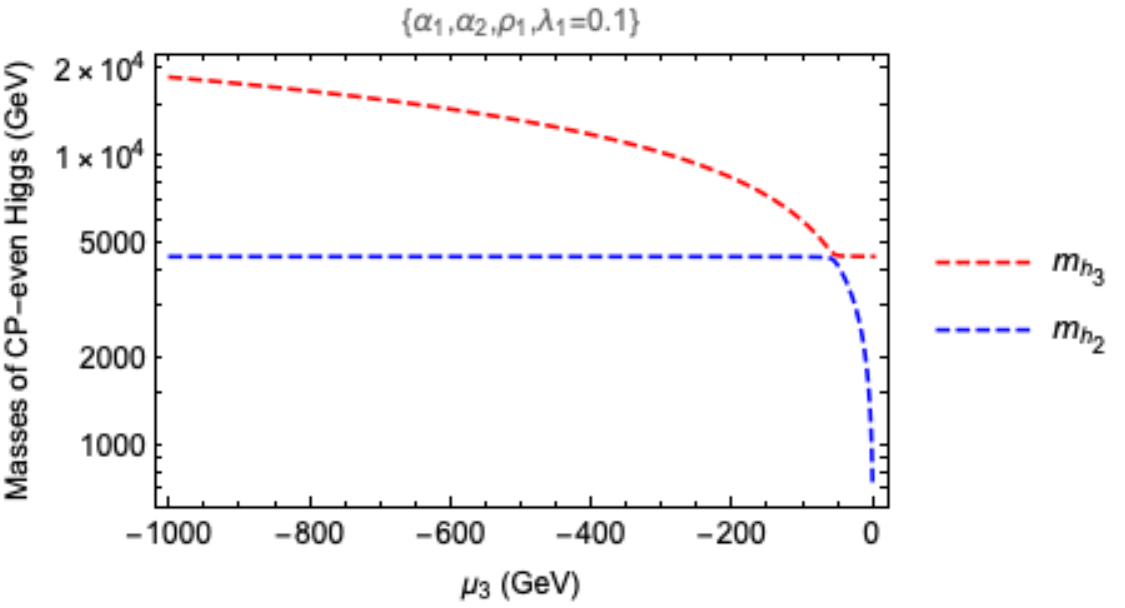}
\caption{Left Panel: CP-even Higgs masses  as  functions of $\mu_3$ for $\alpha_1=\alpha_2=\lambda_1=\rho_1=0.001$. Right panel : CP-even Higgs mass  as  functions of $\mu_3$ for $\alpha_1=\alpha_2=\lambda_1=\rho_1=0.1$.}
\label{fig:evenmu3_dep}
\end{figure}

\begin{figure}[htb!]
\centering
\includegraphics[scale=0.45]{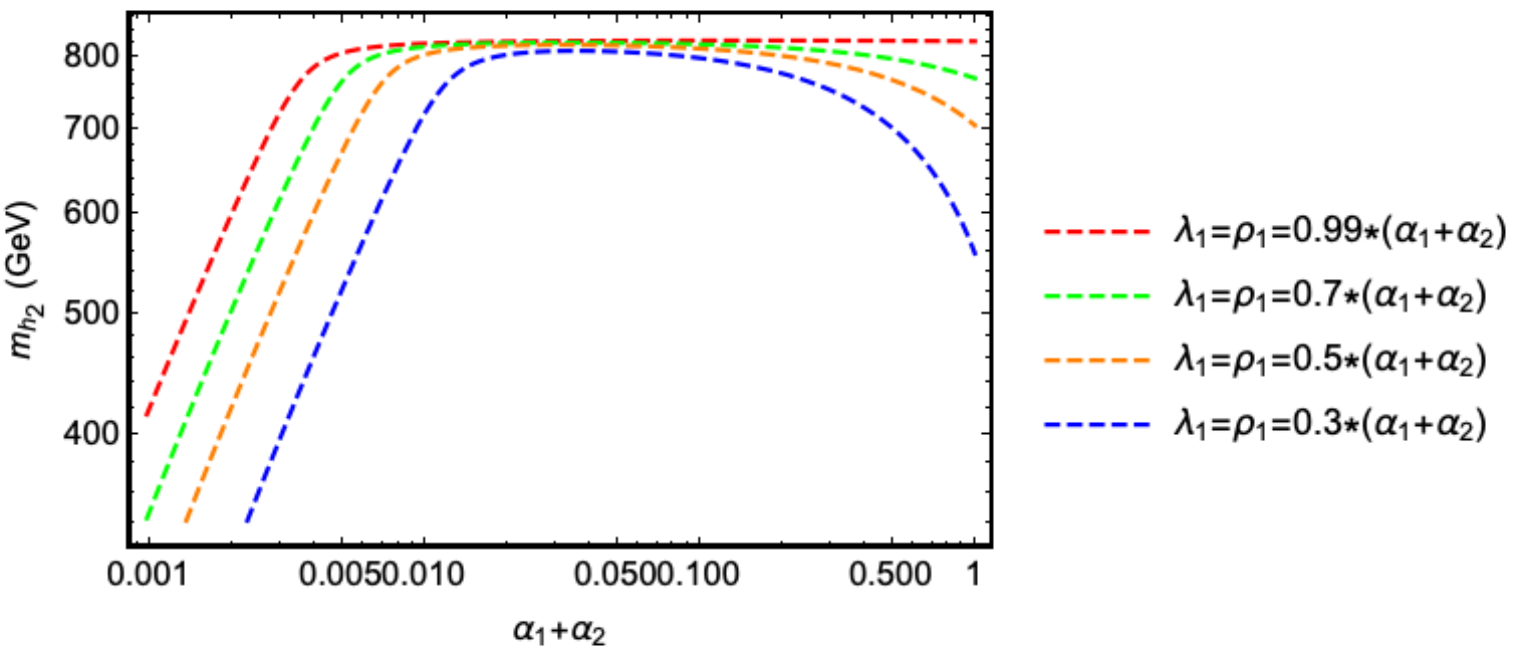} \qquad
\includegraphics[scale=0.45]{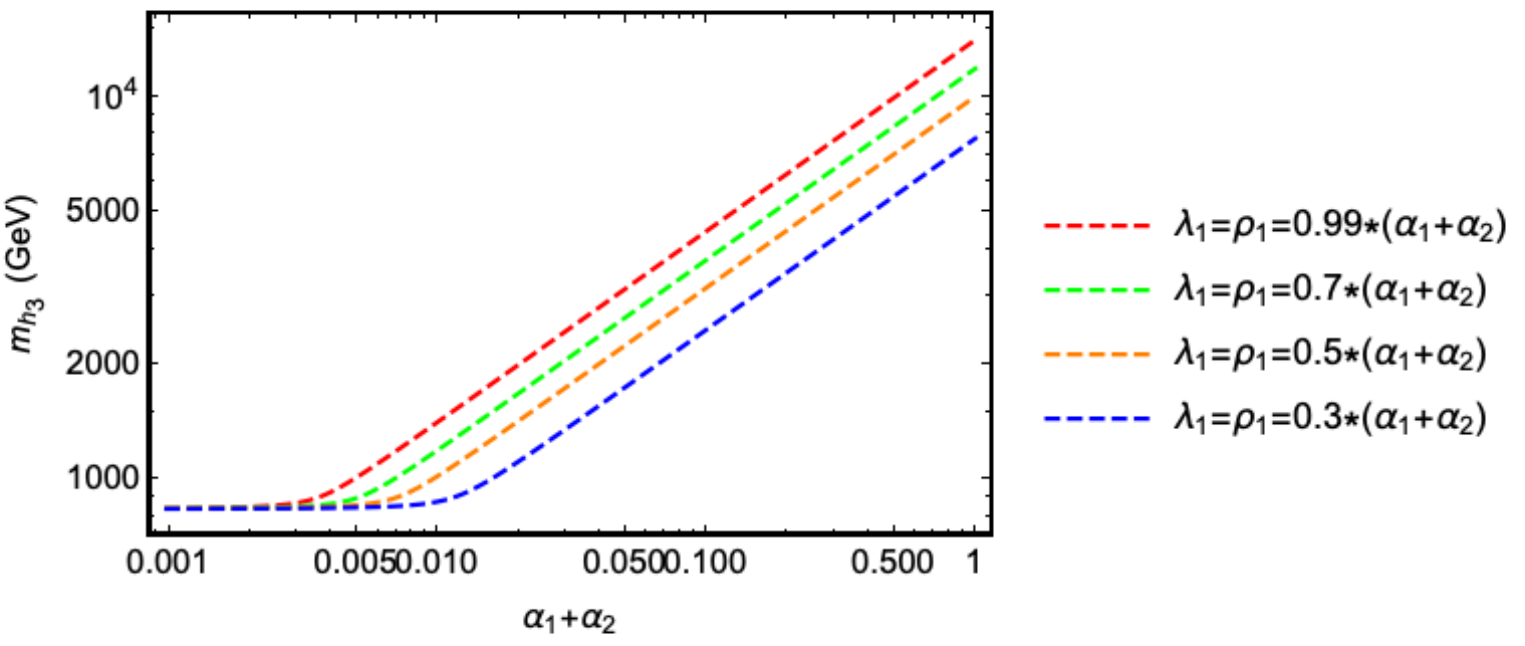}\\
\includegraphics[scale=0.46]{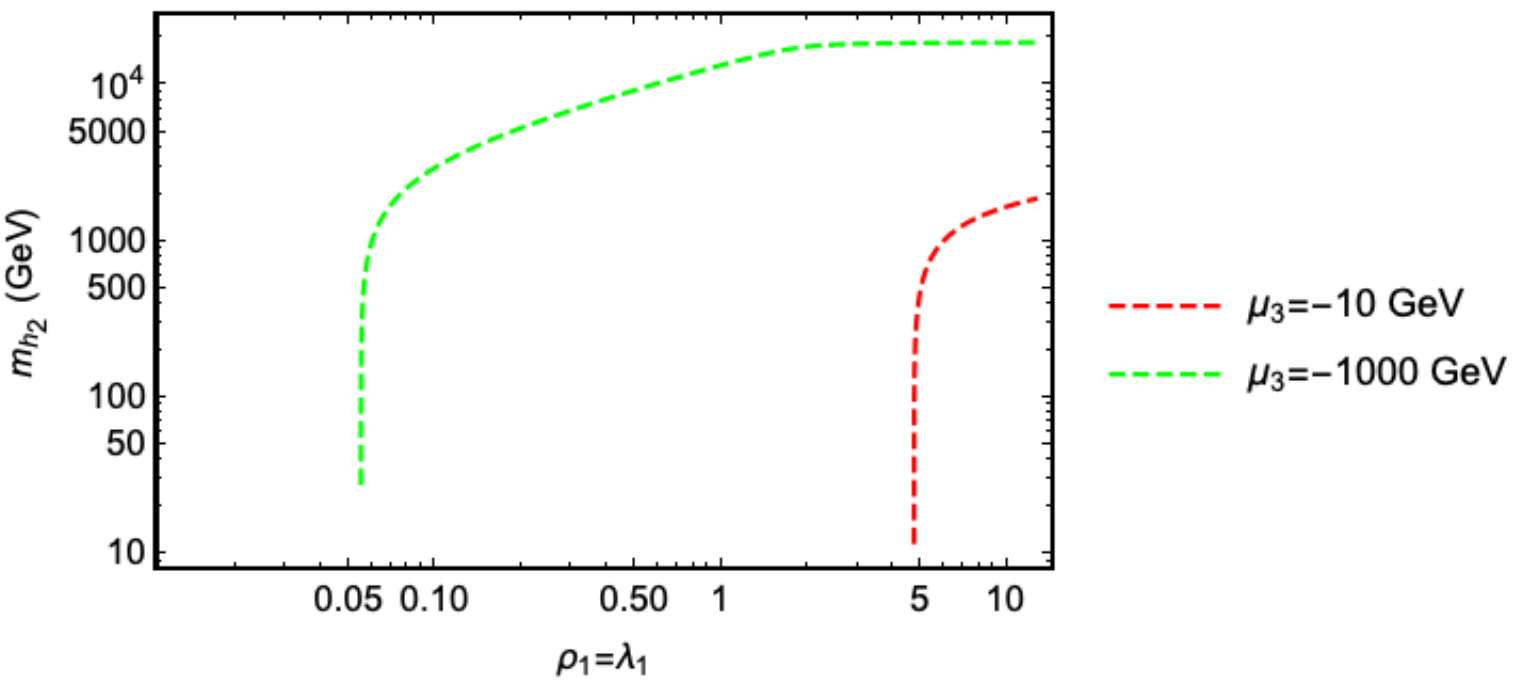}\qquad
\includegraphics[scale=0.46]{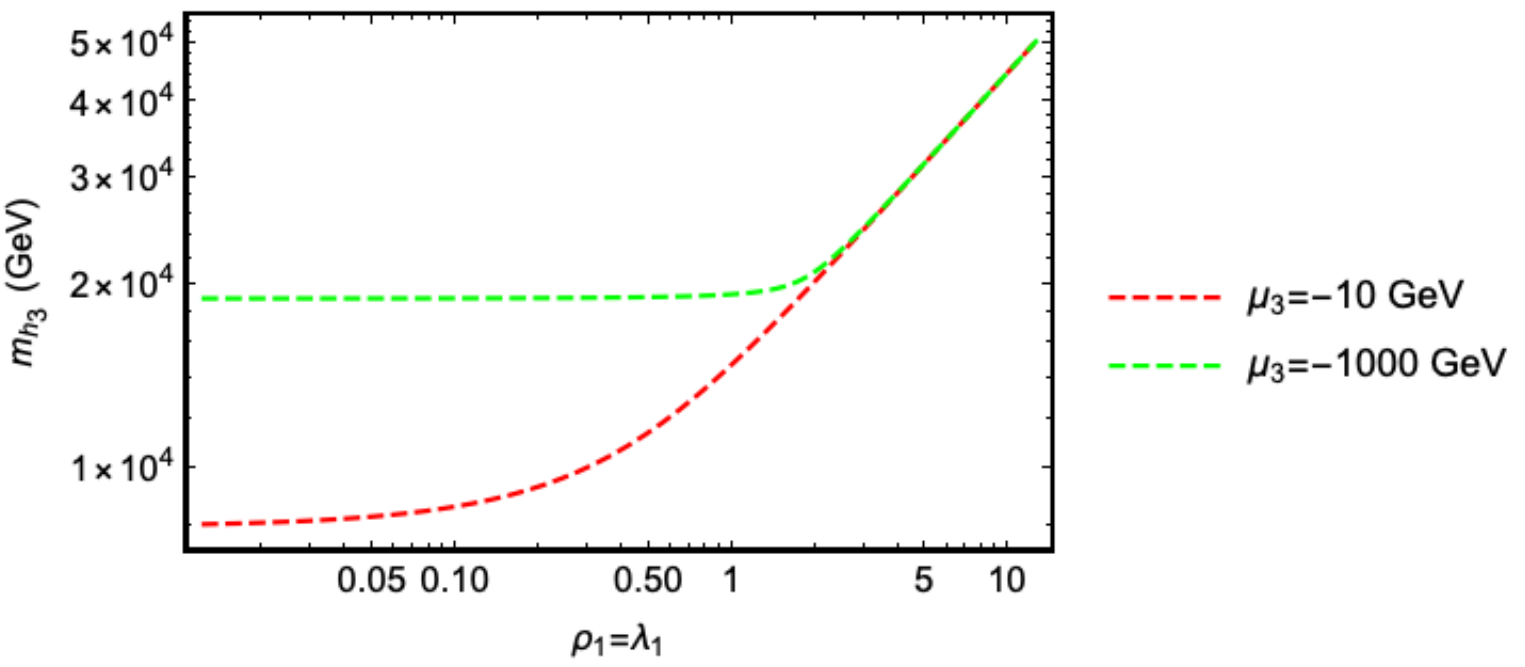}
\caption{Dependence on the Higgs mass on the  couplings in the potential. Top left panel: $h_2$ Higgs mass as a function of $\alpha_1+\alpha_2$ for various values of $\lambda_1\, \rho_1$,  in the lighter mass region;  Top right panel: same as in the left panel, but for $h_3$  Higgs mass; Bottom left panel: $h_2$ Higgs  mass as a function of $\lambda_1=\rho_1$, for two values of $\mu_3$ the heavier mass region;  Bottom right panel: same as in the left panel, but for $h_3$  Higgs mass.}
\label{fig:alpha_dep}
\end{figure}

In the above, we plotted some graphs for the Higgs masses based on choosing some values for the parameters in the potential consistent with vacuum stability. That choice yields some values for the masses. We explore further the implications on the restrictions of the parameters in the scalar potential Eq. \ref{eq:pot} to explore if we can obtain any  restrictions on the Higgs masses.  

In general, it is difficult to impose firm restrictions on Higgs masses from the constraints on scalar couplings, as the Higgs masses depend on several parameters.

 We  investigated first  the dependence  of the pseudoscalar masses, $m_{A_1}$ and $m_{A_2}$ on the parameters $\alpha_2, \, \alpha_3$ and $\mu_3$, which are, in addition to the VEVs, responsible for determining the masses.   In addition to the parameters appearing in the charged Higgs masses, the pseudoscalar masses also depend on $\lambda_2$, which we varied from  $-0.001 \to -1$.  
  
 In plotting Fig. \ref{fig:Higgs},  we considered three values of $v_L$, noting that $v_L$ is constrained from down type quark mass in ALRM:

\begin{enumerate}
\item $v_L \sim 5.95$ GeV (the case with Yukawa coupling  $\sim {\cal O}(1)$),
\item $v_L \sim 1.68$ GeV (minimum value as Yukawa couplings is constrained from perturbativity criteria,
\item $v_L \sim 3.5$ GeV (somewhat arbitrary intermediate case).
\end{enumerate}
 
 There are two competing terms involved in the mass expressions for $m_{A_1}$. If we fix $v_L=5.95$ GeV, then for $\lambda_2= -1.0$,  $|\mu_3|$ must be larger than 700 GeV, otherwise the mass of the pseudoscalar $A_1$ will be imaginary.  Similarly, for $v_L = 3.5\, (1.68)$ GeV,  for $\lambda_2 = -1.0$  values for $|\mu_3|$  are restricted to $|\mu_3 | > 1200\, (2400)$ GeV.  We show the plot of $m_{A_1}$ versus $m_{A_2}$ in Fig. \ref{fig:pseudoscalarHiggs},  for  $v_L=1.68$ GeV (left), and $v_L=5.95$ GeV (right) with no constraints, and  for fixed $\lambda_2=-1$, which yields lower pseudoscalar masses. For the case where $v_L=3.5$ GeV, we verified that the mass range for $m_{A_2}$ lies in the middle of these ranges, and we do not show the plot here.

From these graphs we can easily see that while $m_{A_1}$ is not significantly altered by varying $v_L$ and  imposing $\lambda_2=-1$,   the range allowed for $m_{A_2}$ increases with decreasing $v_L$. For $v_L=5.95$ GeV,  $m_{A_2} >14.5$ TeV, while for $v_L=1.68$ GeV,  $m_{A_2} >51$ TeV.

\begin{figure}[htb!]
\centering
\includegraphics[scale=0.47]{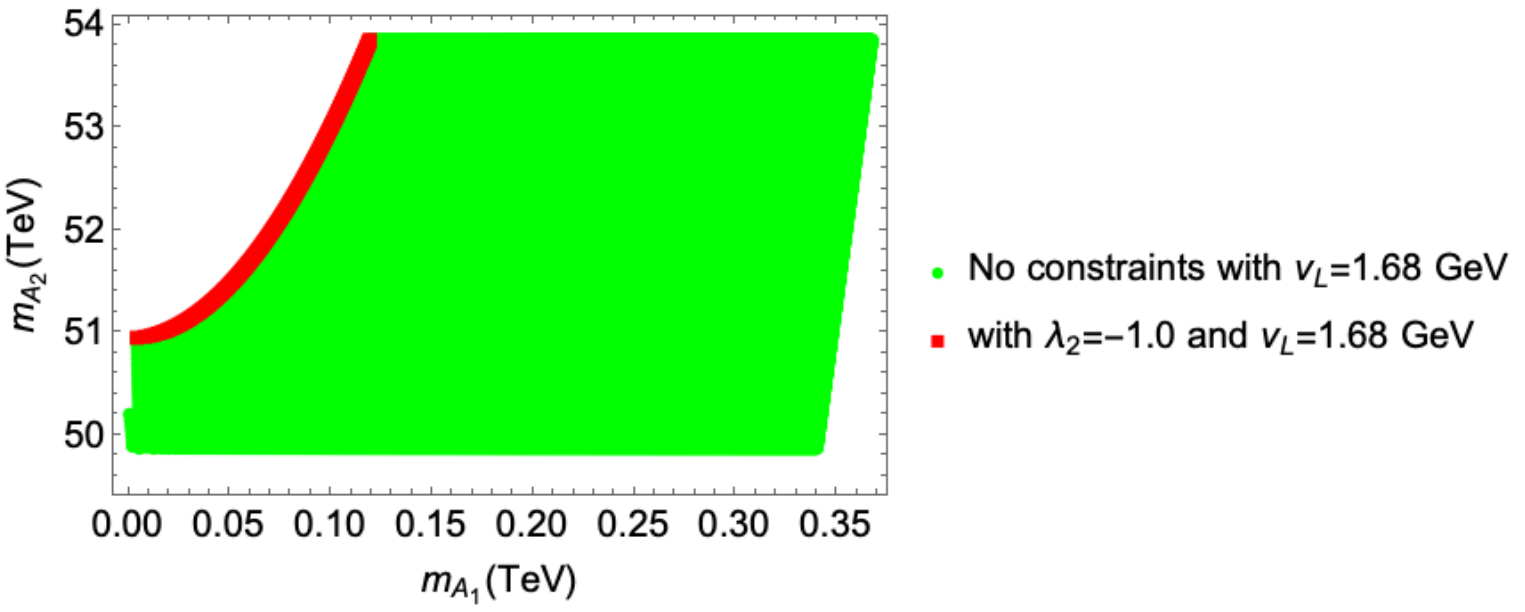} 
\includegraphics[scale=0.47]{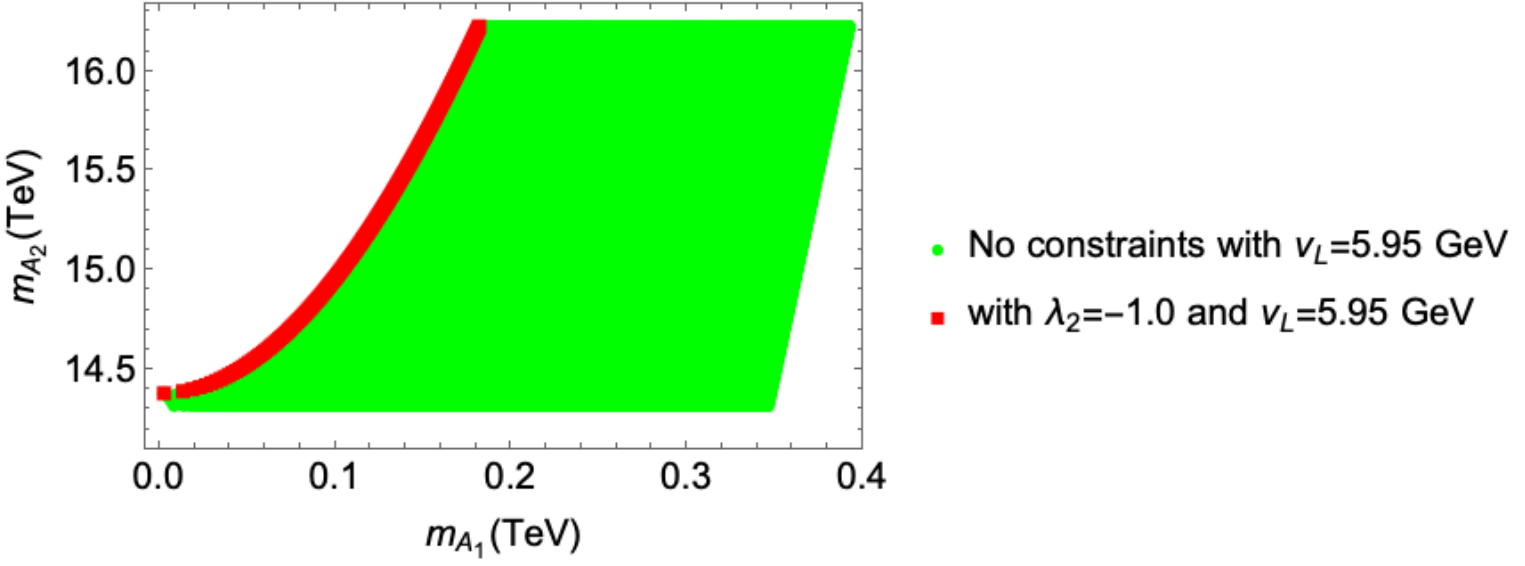}
\caption{Parameter space of the pseudoscalar Higgs masses obtained by varying $\mu_3$, and the dependence  $\lambda_2$  couplings in the scalar potential, and on the VEV $v_L$. Left: $v_L=1.68$ GeV, Right: $v_L=5.95$ GeV.}
\label{fig:chargedHiggs}
\end{figure}

Next we explored the parameter space for the charged Higgs boson masses, $m_{h_1^\pm}$ and $m_{h_2^\pm}$, whose expressions depend on $\alpha_2, \, \alpha_3$ and $\mu_3$ (in addition to the VEVs). We  varied the parameter $\mu_3$  in the region allowed by constraints on the pseudoscalar masses, and  the corresponding $\alpha$'s in the $0.001 \to 1$. 

 We looked at two possibilities:
\begin{itemize}
\item  Case - I: Without the restriction   $\alpha_2 = \alpha_3$ (which we have solely found from our vacuum analysis);
\item  Case - II: Including the restriction $\alpha_2 = \alpha_3$ which restricts the parameter space significantly.
 \end{itemize}
Our results are shown in Fig. \ref{fig:chargedHiggs}.

 From this plot we can see that $m_{h_2^\pm}$ values are restricted to be around 40 GeV  (very low mass) for  $\alpha_2 = \alpha_3$, while $m_{h_1^\pm}$ values are unrestricted and increase with decreasing $v_L$. The restrictions are  $m_{h_1^\pm}> 14$ TeV for $ v_L=5.95$ GeV, while $m_{h_1^\pm}> 50$ TeV for $v_L=1.68$ GeV. These results are consistent with those shown in  Fig. \ref{fig:Higgs}. In addition, we verified that these masses are insensitive to variations in the VEVs $v_R$ and $k$. (Note that $v_R$ is fixed by constraints on $Z_R$ mass.)
  
\begin{figure}[htb!]
\centering
\includegraphics[scale=0.47]{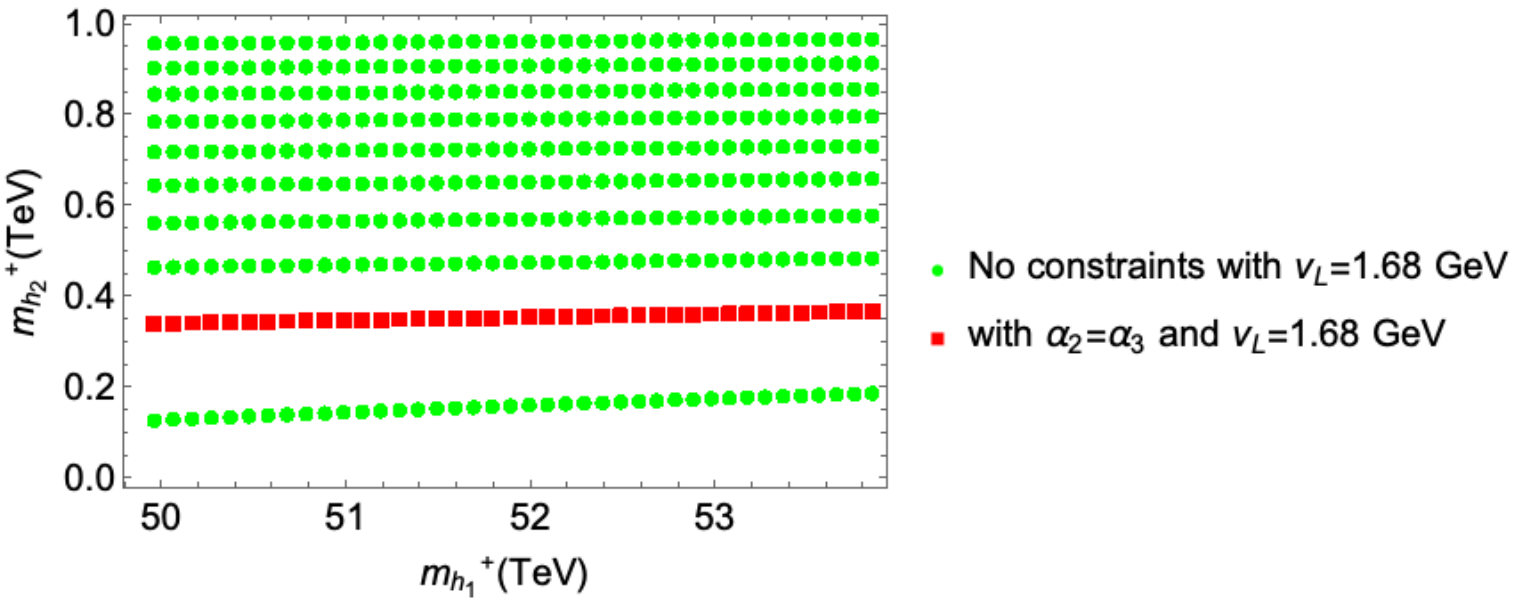} 
\includegraphics[scale=0.47]{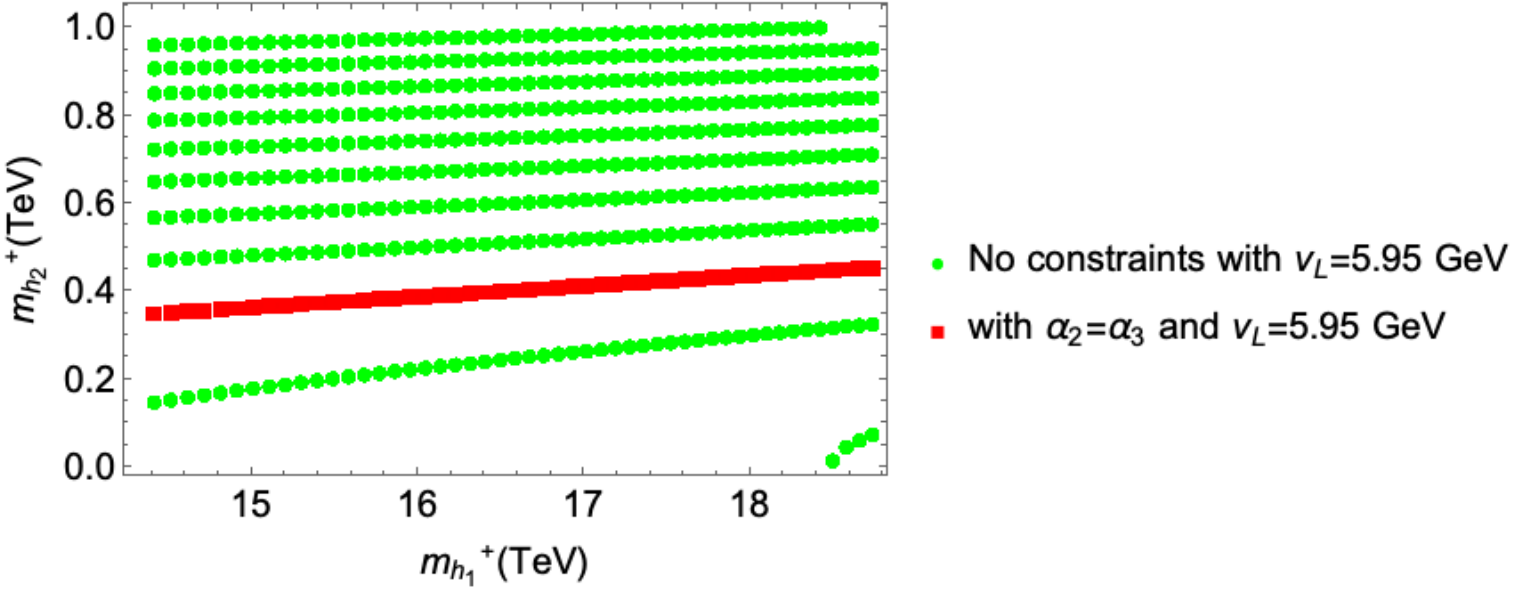}
\caption{Parameter space allowed for the charged Higgs masses on the  VEV $v_L$ while varying $\mu_3$ and $\alpha_2, \, \alpha_3$.  Left: $v_L=1.68$ GeV, Right: $v_L=5.95$ GeV.}
\label{fig:pseudoscalarHiggs}
\end{figure}

Unfortunately, no such simple plots can be obtained for the (CP-even) neutral Higgs masses. Eqs. \ref{even1}-\ref{even2} give the explicit expressions for the masses, which exhibit a complicated dependence of the parameters in the scalar potential. While definite values can be obtained for specific choices of the parameters, we were unable to find more general predictions or restrictions on these masses.

\section{Consistency of  parameters and their constraints}
\label{sec:comments}

In the previous sections we have used several analytical and numerical tools i.e., potential minimization, copositivity requirements, Higgs mass arguments as well as numerical analysis of potential to constrain the parameter spaces spanned by scalar potential parameters. We could instead perform san alternative analysis as in \cite{Ashrythesis}. Using some shorthand redefinition of parameters as, $\alpha_{12} \equiv \alpha_1 + \alpha_2, \alpha_{13} \equiv \alpha_1 + \alpha_3$ and $\lambda_{12} \equiv \lambda_1 + 2\lambda_2$, we observe that we can restate the constraints in terms these combinations only. Copositivity requirements from  our analysis ensure $\lambda_2 \leq 0, \alpha_2 - \alpha_3 \leq 0, \lambda_1, \rho_1 \geq 0$ and $\rho_2 > \rho_1 \geq 0$. In addition from Higgs mass analysis, positivity of scalar masses require $\alpha_2 - \alpha_3 \leq 0$ and $\mu_3 < 0$. Combining these conditions we obtained $\alpha_2 = \alpha_3$ which implies $\alpha_{12}=\alpha_{13}$. While in  \cite{Ashrythesis} one distinguishes 8 possible conditions for vacuum stability, as we have found $\alpha_{12}=\alpha_{13}$,  these conditions are reduced to 4 i.e., when $\alpha_{12}$ and $\alpha_{13}$ have different signs they must be both zero. We can reconsider the 4 different conditions as :
\begin{enumerate}
\item $\alpha_{12}, ~\alpha_{13} \geq 0, ~~\lambda_{12} \geq 0$ .These conditions always obey all the derived copositivity criteria.
\item $\alpha_{12}, ~\alpha_{13} \geq 0, ~\lambda_{12} \leq 0$. The conditions for vacuum stability are : 
\begin{equation}
    \lambda_1 + \lambda_2 \geq 0, ~~\lambda_{12}^2 + 2\lambda_1 \lambda_2 \leq 0.
\end{equation}
Combining $\lambda_1+\lambda_2 \geq 0$ with $\lambda_{12} \leq 0$ and $\lambda_1 \lambda_2 \leq 0$ along with the copositivity condition $\lambda_2 \leq 0$; these imply $\lambda_1 > 0$ which is consistent with the copositivity criteria. The surviving conditions here are:
\begin{equation}
    \lambda_1 + \lambda_2 \geq 0,~~ |\lambda_1 \lambda_2| \geq \frac{\lambda_{12}^2}{2}.
\end{equation}
\item $\alpha_{12}, ~\alpha_{13} \leq 0, ~~\lambda_{12} \geq 0$ .The previously derived conditions are :
\begin{equation}
    \lambda_1 \rho_1 - \alpha_{12}^2 \geq 0, ~~\alpha_{12}^2 (\rho_1 - \rho_2) \geq 0
\end{equation}
The first condition clearly establishes $\lambda_1 \rho_1 \geq 0$, also as $\lambda_{12}^2$ is positive, the second condition ensures that $\rho_1 - \rho_2 \geq 0$ which is in conflict with the copositivity argument $\rho_1 - \rho_2 \leq 0$ unless the surviving condition becomes $\alpha_{12}=\alpha_{13}=0$.
\item $\alpha_{12}, ~\alpha_{13} \leq 0,~~ \lambda_{12} \leq 0$. The stability conditions are : 
\begin{equation}
    \alpha_{12}=\alpha_{13} =0, ~~\lambda_1 + \lambda_2 \geq 0, ~~\lambda_{12}^2+2\lambda_1\lambda_2 \leq 0.
\end{equation}
Combining $\lambda_1+\lambda_2 \geq 0$ with $\lambda_{12} \leq 0$ as well as $\lambda_1 \lambda_2 \leq 0$ with the copositivity criteria $\lambda_2 \leq 0$ imply that $\lambda_1 > 0$ (consistent with our copositivity requirement). Now the surviving conditions are :
\begin{equation}
    \lambda_1 + \lambda_2 \geq 0, ~~\alpha_{12}=\alpha_{13}=0, ~~|\lambda_1\lambda_2| \geq \frac{\lambda_{12}^2}{2}.
\end{equation}
\end{enumerate}
From the four different regions of parameter space we looked at we can summarize: case (i) $\Rightarrow$ no restrictions and cases (ii), (iii) and (iv) (the latter of which combines the restrictions from both (ii) and (iii)). Thus condition (iv) yields most restrictive region for ALRM potential study. Imposing $\alpha_2=\alpha_3$ as well, the number of independent constraints on $\alpha_i$ and $\rho_i$ parameters in Eqs. \ref{1st} and \ref{2nd} becomes 2, the corresponding conditions can be summarized as follows (defining $\alpha_2=\alpha_3=\alpha^\prime$)
\begin{eqnarray}
   (i)~~ \alpha_1 + \alpha^\prime + \sqrt{\lambda_1 \left(\frac{\rho_1+\rho_2}{2}\right)} > 0; ~~~~~~ (ii)~~ \alpha_1 + \alpha^\prime + \sqrt{\lambda_1 \rho_1} > 0
\end{eqnarray}
These relations clearly constraints $\alpha^\prime \equiv \alpha_2 = \alpha_3 \geq 0$. Note also that they are completely consistent with the constraints Eqs. \ref{eq:vac_stability}. This completes and confirms our analysis of vacuum stability in ALRM.

\section{Conclusion}
\label{sec:conclusion}
In this work we analysed the vacuum structure of the alternate left-right model ALRM,  which is another possible option to breaking $E_6$ grand unified group into $SU(2)_L \times SU(2)_R$. The advantage of the model over the more commonly studied left right symmetric model is the absence of FCNC in the Higgs sector. Thus the Higgs states in this model can be light and within reach of the LHC.  With this aim  we have explored the vacuum structure of the ALRM scalar sector with symmetry breaking implemented using Higgs bidoublet $\Phi$ and doublets $\chi_{L,R}$. We have  use the  ``bounded from below" and copositivity criteria to obtain stable vacuum structure. We performed an analytical study and numerical investigations of the ALRM potential with these criteria, thereby restricting the parameter space.  Possible gauge equivalent vacua that may lead to charge-breaking are identified, however they are seen not to enter our later analysis. 

These model building constraints are then combined with mass restrictions on the Higgs states, some of which decay into exotic fermions and are long-lived, while some others can be produced through Drell-Yan processes and whose masses can be restricted from Higgs searches.  We also showed the dependence of the CP-even, CP-odd and charged Higgs masses on the various parameters of the model and gave some examples of light and heavier masses variations. Based on restrictions on parameters in the scalar potential, one pseudoscalar Higgs and one charged Higgs state is very light (can be less that 100 GeV) while the others lie in the TeV range.

Finally, we  compared our analysis to a previous investigation of the structure of the vacuum, and showed consistency within their parameter space with their results. Our analysis shows that the parameter space of ALRM is much more restricted than previously explored, and that Higgs masses depend on a relatively small number of parameters and that the masses of scalars, pseudo-scalars and charged Higgs are heavily interconnected, rendering the model very predictive.

\begin{acknowledgments}
The work of MF  has been partly supported by NSERC through grant number SAP105354. SS is thankful to UGC and IRCC, IIT Bombay for financial support in her research work.
\end{acknowledgments}
\bibliographystyle{JHEP}
\bibliography{ALRM_vac}
\end{document}